\documentclass[prd,twocolumn,tightenlines,preprintnumbers,showpacs,superscriptaddress,notitlepage,nofootinbib,eqsecnum,floatfix,longbibliography,aps,10pt]{revtex4-1}
\pdfoutput=1
\usepackage{CJK}
\usepackage[utf8]{inputenc}
\usepackage[T1]{fontenc}
\usepackage{amsmath}
\usepackage{mathtools}
\usepackage{amsfonts,dsfont}
\usepackage{bm,bbm}
\usepackage{graphicx}
\usepackage{xcolor}
\usepackage{enumitem}

\usepackage{hyperref}
\hypersetup{
    colorlinks=true,     % false: boxed links; true: colored links
    linkcolor=blue,      % color of internal links
    citecolor=blue,      % color of links to bibliography
    filecolor=blue,      % color of file links
    urlcolor=blue        % color of external links
}

\makeatletter
\def\l@subsubsection#1#2{}
\makeatother

\begin{document}
\title{Integer Programming from Quantum Annealing and Open Quantum Systems}

\author{Chia~Cheng~Chang}
\affiliation{RIKEN iTHEMS, Wako, Saitama 351-0198, Japan}
\affiliation{Department of Physics, University of California, Berkeley, California 94720, USA}
\affiliation{Nuclear Science Division, Lawrence Berkeley National Laboratory, Berkeley, California 94720, USA}
	\author{Chih-Chieh~Chen }
\affiliation{R\&D Group, Grid Inc., Tokyo 107-0061, Japan}
\affiliation{Department of Physics, University of California, Berkeley, California 94720, USA}
\author{Christopher K\"orber}
\affiliation{Institut f\"ur Theoretische Physik II, Ruhr-Universit\"at Bochum, D-44780 Bochum, Germany}
\affiliation{Department of Physics, University of California, Berkeley, California 94720, USA}
\affiliation{Nuclear Science Division, Lawrence Berkeley National Laboratory, Berkeley, California 94720, USA}
\author{Travis~S.~Humble}
\affiliation{Computational Sciences and Engineering, Oak Ridge National Laboratory, Oak Ridge, Tennessee, 37831, USA}
\author{Jim~Ostrowski}
\affiliation{Industrial and Systems Engineering, University of Tennessee, Knoxville, Tennessee 37996, USA}

\newcommand{\alert}[1]{\textbf{\color{red}{#1}}}
\renewcommand{\vec}[1]{\boldsymbol{#1}}

\newcommand{\ghissue}[2]{
 \noindent\fbox{\parbox{0.49\textwidth}{
   \alert{[#1]}%
   \\%
   \href{https://github.com/cchang5/quantum\_linear\_programming/pull/#2}{See GitHub issue #2}}%
 }
}

\begin{abstract}
 While quantum computing proposes promising solutions to computational problems not accessible with classical approaches, due to current hardware constraints, most quantum algorithms are not yet capable of computing systems of practical relevance, and classical counterparts outperform them.
 To practically benefit from quantum architecture, one has to identify problems and algorithms with favorable scaling and improve on corresponding limitations depending on available hardware.
 For this reason, we developed an algorithm that solves integer linear programming problems, a classically NP-hard problem, on a quantum annealer, and investigated problem and hardware-specific limitations.
 This work presents the formalism of how to map ILP problems to the annealing architectures, how to systematically improve computations utilizing optimized anneal schedules, and models the anneal process through a simulation.
 It illustrates the effects of decoherence and many body localization for the minimum dominating set problem, and compares annealing results against numerical simulations of the quantum architecture.
 We find that the algorithm outperforms random guessing but is limited to small problems and that annealing schedules can be adjusted to reduce the effects of decoherence.
 Simulations qualitatively reproduce algorithmic improvements of the modified annealing schedule, suggesting the improvements have origins from quantum effects.
\end{abstract}

\preprint{RIKEN-iTHEMS-Report-20}

\maketitle
%%%%%%%%%%%%%%%%%%%%%%%%%%%%%%%%%
%%%%%%%%%%%%%%%%%%%%%%%%%%%%%%%%%
%%%%%%%%%%%%%%%%%%%%%%%%%%%%%%%%%
%\tableofcontents

\flushbottom
\maketitle

%========================================================================================
\section{INTRODUCTION}
\label{sec:introduction}
%========================================================================================
Integer Linear Programming (ILP) is an integer optimization problem subject to inequality constraints
\begin{align}
 \label{eq:initial-ip-def}
 \vec x_0 = \mathrm{arg} & \min\limits_{x}\left(\sum_i c_i x_i\right)
\end{align}
subject to
\begin{align}
 \label{eq:ilp-constraints}
  & \sum_i A_{ai}x_i +b_a \leq 0 \,, \\
  & x_i  \in \mathbbm{Z} \geq 0\, ,
\end{align}
where $i=1, \cdots,  N$ is the number of dependent variables and $a=1, \cdots, M$ the number of constraint equations.

ILP is a commonly tackled problem applicable to situations such as scheduling, network optimization, and graph optimization such as the minimum dominating set problem (MDS).
In general, ILP is classically NP-complete, and as a result, heuristic methods are employed~\cite{GLOVER1986533, doi:10.1287/ijoc.1.3.190, doi:10.1287/ijoc.2.1.4}.
Standard classical heuristic algorithms follow a greedy scheme which iteratively approximates optimal solutions--starting from a random initial guess, these algorithms apply locally optimal choices at each step.
The NP-hardness of ILP can be understood by realizing that while the solution to an $n$-dimensional linear problem must lie on the vertices of the feasibility region, while the optimal integer solution may in general be at any integer solution inside the feasibility region.
While greedy algorithms do not guarantee optimal global solutions, they find approximate solutions in polynomial time, which can be utilized in further computations.

\begin{figure*}
	\centering
	\begin{tabular}{p{0.2\textwidth}p{0.1\textwidth}p{0.2\textwidth}p{0.1\textwidth}p{0.2\textwidth}}
	\includegraphics[width=0.2\textwidth]{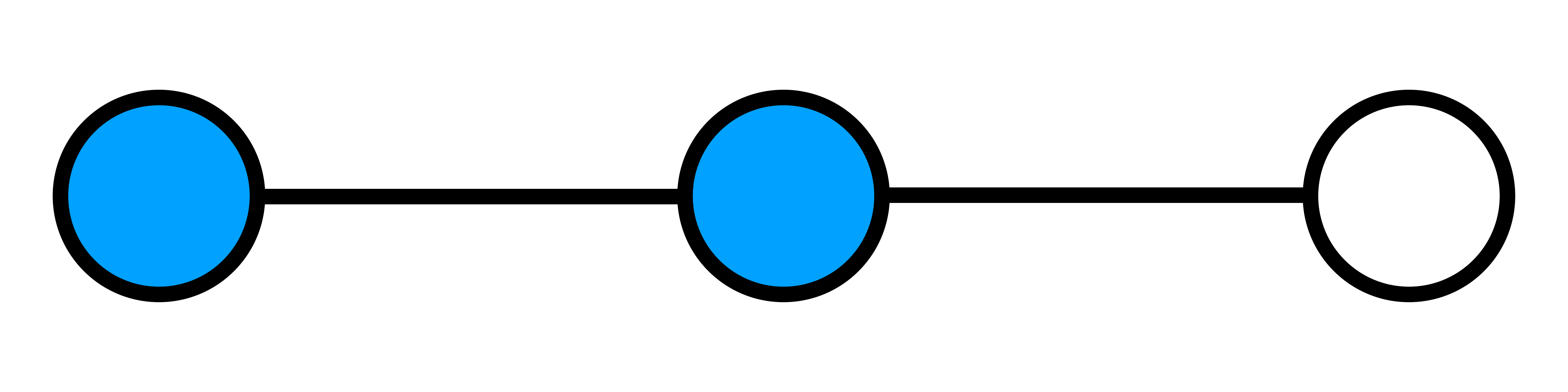}
&&
	\includegraphics[width=0.2\textwidth]{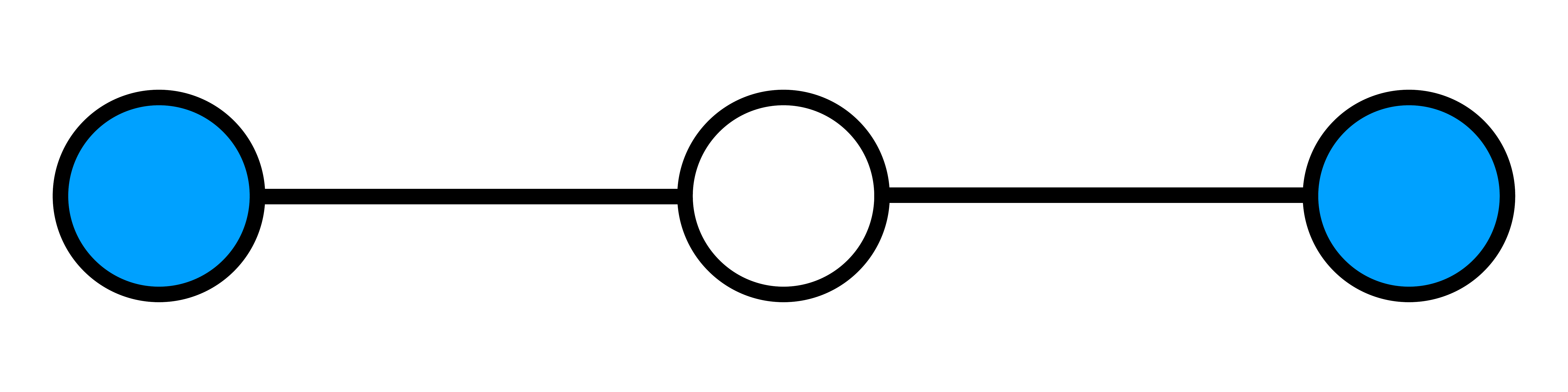}
&&
	\includegraphics[width=0.2\textwidth]{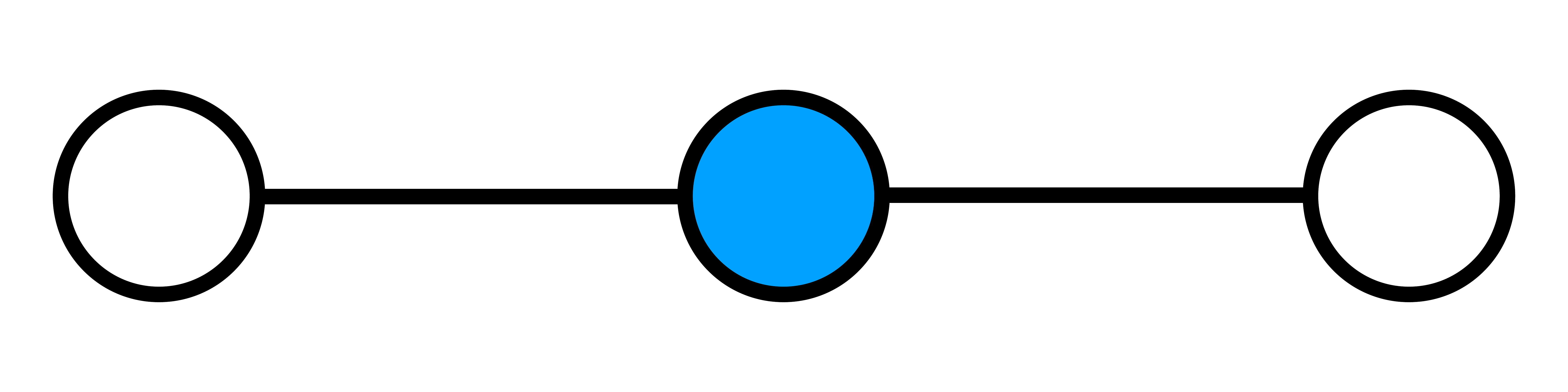}\\
	\centering\textbf{a} && \centering\textbf{b} && \centering\textbf{c}
	\end{tabular}
	\caption{Example of different dominating sets for $G(V, E)$. Vertices in the dominating set $D$ are highlighted in blue. {\textbf{a)}} A dominating set of $G$ with domination number $\overline{\overline{D}} = 2$. {\textbf{b)}} A minimal dominating set of $G$ with domination number of $\overline{\overline{D}} = 2$. {\textbf{c)}} The MDS of $G$ with domination number of $\overline{\overline{D}} = 1$.}
	\label{fig:dominating_sets}
\end{figure*}

An important ILP application is the MDS problem, which is representatively considered in this work.
For a given a graph $G(E,V)$, defined by the set of $V$ vertices and $E$ edges, a dominating set $D$ is a specific subset of vertices $D \subseteq V$.
In particular, $D$ is a dominating set if all vertices in $V$ but not in $D$ are adjacent to at least one vertex in $D$.
This is equivalent to requiring the set of nearest-neighbor vertices of $D$ (exclusive) and $D$ cover all vertices $N(D) \cup D = V$ (an example is given by Fig.~\ref{fig:dominating_sets}a).
The set $D$ is a minimal dominating set if there is no proper subset of $D$ that is a dominating set, {\it{i.e.}}, the removal of any vertex in $D$ results in $N(D) \cup D  \neq V$.
An example is given by Fig.~\ref{fig:dominating_sets}b.
The domination number of $D$ is given by the cardinality of $|D| \equiv \overline{\overline{D}}$.
The MDS is defined by $D$ with the smallest domination number.
Fig.~\ref{fig:dominating_sets}c shows an example of the MDS of $G(V, E)$ and is different from the minimal dominating set.
We emphasize that while the maximum independent set is always a minimal dominating set as exemplified by Fig.~\ref{fig:dominating_sets}b, the MDS, in general, can have a smaller domination number.
As a result, the solution to the dominating set problem can not be obtained by solving the maximum independent set problem, a well-studied problem for quantum annealers.

For general graphs, existing algorithms on classical computers find MDS solutions in exponential time $\sim O( 1.5^n)$ \cite{Fomin2009, vanRooij2009} or approximate solutions in polynomial time. For example, greedy algorithms locally optimize decisions about which nodes to add to the dominating set.
Thus one is guaranteed to find a dominating set but not necessarily an MDS.

In this work, we present a method to obtain optimal solutions to ILP by employing quantum annealing methods.

Current implementations of quantum annealing solve the quadratic binary optimization problem (QUBO) by slowly varying a time-dependent Hamiltonian~\cite{1998PhRvE..58.5355K, 2000quant.ph..1106F, RevModPhys.80.1061}.
Through the adiabatic theorem of quantum mechanics, the annealer is initially prepared in a trivial ground state while the final Hamiltonian encodes the solution to the ILP.
Due to the explosion in research efforts towards hardware implementations of quantum annealers and future improvements to the annealing schedule~\cite{doi:10.7566/JPSJ.89.044001}, mapping ILP to QUBO provides a path forward towards obtaining optimal solutions to the class of integer optimization problems~\cite{2018Glover}.

In the simplest model for quantum annealing, the pure state of a quantum system evolves adiabatically to prepare an eigenstate of the encoded problem Hamiltonian. The time-dependent Hamiltonian is given as
\begin{equation}
 H(t) = A(t) H^{\textrm{init}} + B(t) H^{\textrm{problem}}, \label{eq:tdhamiltonian}
\end{equation}
and $H^\textrm{init}=-\sum_i\sigma^x_i$ (on the D-Wave), while $H^\textrm{problem}$ encodes the problem to be solved. In practice, and as discussed in Sec.~\ref{sec:methods:lindblad}, modeling dynamics that arise during quantum annealing requires a more robust description of the thermally populated mixed quantum states and the open dynamical processes that govern population of the sought-after eigenstate.

The mapping of ILPs to QUBOs we propose is realized by introducing slack variables $s_a$ which turn the inequalities Eq.~\eqref{eq:ilp-constraints} to equalities
\begin{align}
  \label{eq:ilp:slack}
  & \sum_i A_{a i}x_i + s_a + b_a = 0, \\
  & x_i, s_a \in \mathbb{Z} \geq 0.
\end{align}
While, in general, the coefficients of the inequality constraints are not required to be integer valued, this real valued inequalities can be trivially rescaled such that $s_i \in \mathbbm{Z}$ given fixed precision coefficients $A_{ij}$ and $b_i$.
Furthermore, this formalism can be generalized to constrained quadratic optimizations \ref{sec:methods:ilp:quadratic}.

We improve the quantum annealer's performance by utilizing annealing offsets, which effectively delay the annealing schedule on a per-qubit basis~\cite{PhysRevA.96.042322,hsu2018quantum,10.1007/978-3-030-14082-3_14}.
Converting to the Ising representation of the the problem Hamiltonian,
\begin{equation}
    \label{eq:HIsing}
     H^{\textrm{problem}} \leftrightarrow H^{\textrm{Ising}} = \sum_{ij} J_{ij} \sigma^z_i \sigma^z_j + \sum_i h_i \sigma^z_i ,
\end{equation}
we recognize that Eq.~(\ref{eq:HIsing}) exhibits spin-glass properties. More specifically, if the $h_i$ coefficients are randomly drawn from a Bernoulli distribution, one expects spin-localization behavior to influence the outcome of the anneal.
In the case of quantum annealing, when an algorithm is mapped to its Ising representation, the values of $h_i$ will frequently take on various values, mimicking spin-glass like behavior.
More explicitly, the spin-glass enters a glassy state when more disorder is introduced ($|h_i|$ becomes large), and as a consequence, the wavefunction experiences many body localization (MBL) effects, the many-body analog of Anderson localization~\cite{doi:10.1146/annurev-conmatphys-031214-014726,PhysRevE.90.022103,RevModPhys.91.021001,ALET2018498,PhysRevB.82.174411,PhysRevLett.109.017202}.

Our improvement strategy is motivated by the MBL hypothesis.
As a result, we employ a modified annealing schedule that relies on partitioning the Hamiltonian into regions of relatively weak and strong external magnetic fields.

To understand which phenomena are relevant for describing the proposed offset study's scaling behavior, whether they are rooted in the quantum nature or related to hardware constraints, we simulate the anneal for a small MDS problem. This simulation solves the von Neumann equations accounting for different quantum decoherence models and explores whether algorithmic improvements on hardware are present in idealized systems.

%========================================================================================
\section{RESULTS}
\label{sec:results}
%========================================================================================

We first present the QUBO mapping for ILP (\ref{sec:results:qa1}), and demonstrate the methodology on an example implementation in case of the Minimal Dominating Set problem (\ref{sec:results:qa}) on the D-Wave quantum annealer. Results from quantum annealing are compared and discussed in contrast to simulations (\ref{sec:results:simulation}).

%----------------------------------------------------------------------------------------
\subsection{QUBO Formulation of ILP}
\label{sec:results:qa1}
%----------------------------------------------------------------------------------------
Following Eq.~(\ref{eq:ilp:slack}), the ILP problems simplifies to solving a system of linear equations on integer valued variables. We map the integer variables $\vec z = \vec x, \vec s$ appearing in Eq.~\eqref{eq:ilp:slack} to qubits under the following transformation~\cite{Chang:2018uoc}
\begin{align}
 z_i = & \sum_{r=0}^{R_i-1} 2^r \psi_{ri}
 \label{eq:int_to_bin}
\end{align}
where $\psi_{ri} \in \{0, 1\}$. The number of qubits used to represent the $i$-th integer variable is allowed to vary with $R_i$.
The integer-vector qubit transformation can be expressed as a rectangular matrix.
For example, a vector of two integer variables $z_0$ and $z_1$ represented by one and two qubits respectively is given as
\begin{align}
 \begin{pmatrix}
  z_0 \\
  z_1
 \end{pmatrix}
 = &
 \begin{pmatrix}
  2^0 & 0   & 0   \\
  0   & 2^0 & 2^1
 \end{pmatrix}
 \begin{pmatrix}
  \psi_{00} \\
  \psi_{01} \\
  \psi_{11}
 \end{pmatrix}
 \equiv T^z \begin{pmatrix}
  \psi_{00} \\
  \psi_{01} \\
  \psi_{11}
 \end{pmatrix}
\end{align}
If all variables are represented by the same number of qubits--$R_i$ is a constant for all $i$--then one can express the transformation as tensor product of bit vectors
\begin{align}
 \mathcal{R} =  & \begin{pmatrix} 2^0 & \dots & 2^{R-1}\end{pmatrix},    \\
 \mathcal{Z} =  & \begin{pmatrix} z_0 & \dots & z_{N-1}\end{pmatrix},    \\
 |\mathds{1}| = & |\mathcal{Z}|,                 \\
 T^z =          & \mathds{1}\otimes \mathcal{R}.
\end{align}

As a result, the $\vec x$ and $\vec s$ map to the binary vector $\vec \Psi$ under the transformation
\begin{equation}
    \label{eq:bit-vector-mapping}
    \begin{pmatrix}
        \vec x \\ \vec s
    \end{pmatrix}
    =
    \begin{pmatrix}
        T_x & 0 \\ 0 & T_s
    \end{pmatrix}
    \begin{pmatrix}
        \vec \Psi_x \\ \vec \Psi_s
    \end{pmatrix}
    =
    T
    \vec \Psi
\end{equation}

The integer linear optimization problem is then solved through the minimization of the quadratic objective function
\begin{align}
 \label{eq:ilp-slack-bit-energy}
 \chi^2(\vec \Psi)
 &=
 \vec c^T T_x \vec \Psi_x + p \left\| A T_x \vec \Psi_x + T_s \vec \Psi_s + \vec b \right\|^2
\end{align}
where $p$ is an external parameter representing the strength of the penalty when violating constraints.
This parameter needs to be sufficiently large, e.g., $p \geq \vec c \cdot \vec x$, to ensure the constraints are indeed fulfilled\footnote{Depending on the problem, $p$ can be smaller as well.
For example, in the case of the MDS, $p\geq 1$ suffices.}.
The objective function $\chi^2$ can be represented as a QUBO Hamiltonian
\begin{align}
 \chi^2(\vec \Psi) =    &
 \vec \Psi^T
 \begin{pmatrix}
  Q_{xx} & Q_{xs} \\
  Q_{sx} & Q_{ss}
 \end{pmatrix}
 \vec \Psi + p\left \| \vec b \right\|^2, \\
 \equiv &  \vec \Psi^T Q  \vec \Psi + C,
 \label{eq:matrix_form}
\end{align}
where
{\small
\begin{align}
 \label{eq:qubo:components}
 Q_{xx} = & p {T_{x}}^T \left[ A^T A + \mathrm{diag} \left(A^T \vec b + \vec b^T A\right) \right] T_x + \mathrm{diag}(\vec c) T_x,                                                                    \\
 Q_{xs} = & Q_{sx}^T = p {T_{x}}^T A^T T_s,                                                                     \\
 Q_{ss} = & p\left[ {T_{s}}^T T_s + \mathrm{diag}\left( {T_{s}}^T \vec b + \vec b^T T_s\right) \right].
\end{align}}
The function $\mathrm{diag}(\vec v)$ transforms a vector $\vec v$ into a diagonal matrix, and absorbs the linear contributions of the QUBO into the diagonal elements of the quadratic representation.

The integer solution to the original ILP is computed by $\vec x^{(0)} = T_x \vec \Psi_x^{(0)}$ and the original solution to the problem is computed by shifting the annealer extracted energy $E \equiv  \chi^2(\vec \Psi^{(0)}) $ by $p \left \| \vec b \right\|^2$.
The slack components of this vector $\vec \Psi_s$ can be utilized to check if the constraints are indeed fulfilled.

%----------------------------------------------------------------------------------------
\subsection{Annealer Results for the Dominating Set}
\label{sec:results:qa}
%----------------------------------------------------------------------------------------
\begin{figure}[b]
	\centering
	\begin{tabular}{cc}
	$G(n):$ &
	\raisebox{-.4\height}{\includegraphics[width=0.8\columnwidth]{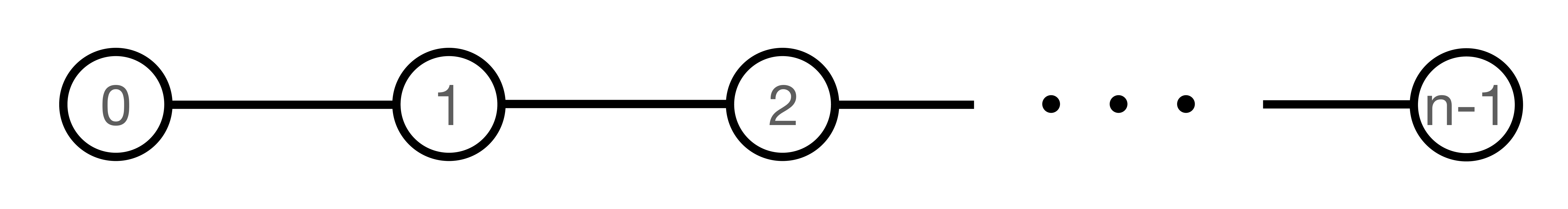}}
	\end{tabular}
	\caption{Linear graphs $G(n)$ used in this study. Nodes denote vertices of the graphs and lines are undirected edges.}
\label{fig:linear}
\end{figure}

We demonstrate the proposed algorithm in order to obtain the MDS on a series of linear graphs $G(n)$, as shown in Fig.~\ref{fig:linear}. This type of graph is chosen because the small number of nearest-neighbor connections is more efficiently embedded into the chimera graph, allowing scaling plots to be generated when using a D-Wave quantum annealer. In particular, the number of qubits required to solve MDS scales at worse as $n_V \log_2 n_V$ where $n_V$ is the number of vertices for a generic graph before minor embedding. Details of the mapping of ILP to MDS is given in Sec.~\ref{sec:methods:mds-qubo}.

For the graph $G(n)$ the MDS solution is known analytically, and contains both unique and degenerate solutions. In particular, the domination number for $G(n)$ is $\lceil n/3 \rceil$ while the number of MDS solutions for $n$ vertices is
\begin{align}
&1 &&\textrm{if} && n\textrm{ mod }3=0,\nonumber \\
&2\lfloor n/3 \rfloor + 1 && \textrm{if}&& n\textrm{ mod }3=1,\nonumber \\
&\lfloor n/3 \rfloor + 2 && \textrm{if} && n \textrm{ mod }3 = 2,
\end{align}
and gives the probability of randomly guessing the MDS of $G(n)$.

\begin{figure}
	\centering
	\includegraphics[width=\columnwidth]{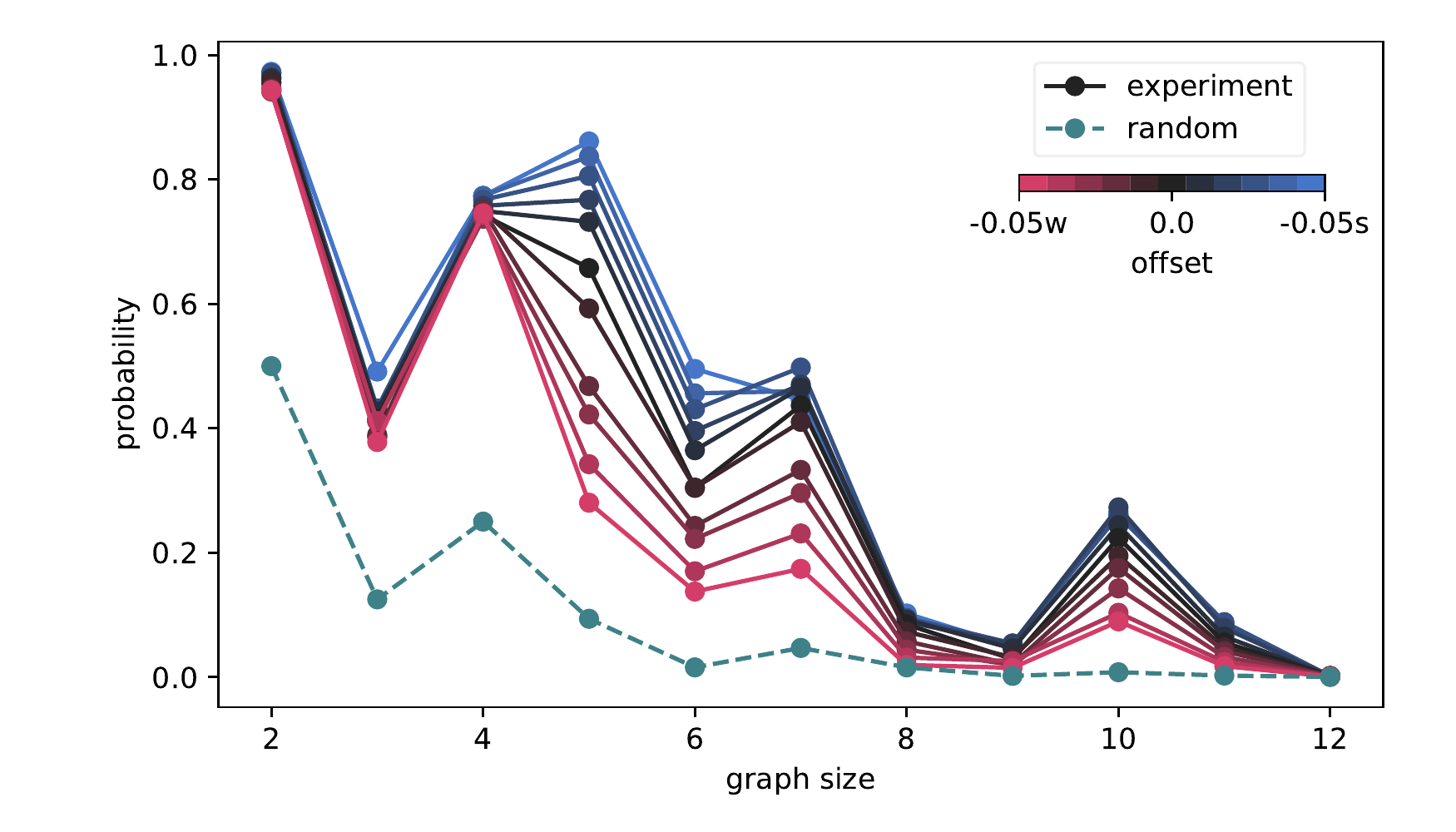}
	\caption{Baseline result of D-Wave (black) compared to random guessing (dashed green). The jagged nature of random guessing reflects the degeneracy of the ground state. Negative offsets with a `s' tag (blue) are results from delays of large values of $|h_i|$, while negative offsets with the `w' label (red) delay the schedule of qubits with small values of $h_i$.}
	\label{fig:baseline}
\end{figure}

For even the smallest graph $G(2)$, the MDS problem is not native to the chimera graph and must be embedded. Following the hypothesis of MBL, we, therefore, must look at the values of $h_i$ after embedding. The qubits split into two groups depending on the value of $h_i$ relative to $(\textrm{max}|\{h\}| + \textrm{min}|\{h\}|) / 2$ given the set of external magnetic fields $\{h\}$ defined by a specific embedding. Further detail is given in Sec.~\ref{sec:methods:minor_embedding}. We study the effects of delaying the anneal schedule of one group of qubits over the other and present the results of this study is shown in Fig.~\ref{fig:baseline}.

Due to near-term limitations, hardware realizations of quantum annealing are unique, and possess for example, different lattice layouts (due to faulty qubits), annealing schedules, and qubit fidelity. For the following studies, we perform experiments specifically on the \texttt{DW\_2000Q\_6} solver. The annealing time is set to 500$\mu$s after performing a study on various annealing times the $G(6)$ graph. The black line (offset$=0.0$) in Fig.~\ref{fig:baseline} shows results from the baseline experiment, without modification to the D-Wave annealing schedule, and observe improvement over random guessing (dashed green).

We explore one avenue towards improving the experiment results by introducing per-qubit annealing offsets into the time evolution.
The blue (red) results delay qubits' annealing subject to stronger (weaker) external fields. We observe improvement (diminishment) in the experimental results when qubits subject to stronger (weaker) final external fields are delayed in the annealing schedule, in agreement with the MBL hypothesis. The phenomena is observed across different problem sizes and hints at the possibility of a generic improvement strategy.

%----------------------------------------------------------------------------------------
\subsection{Simulation Results}
\begin{figure}[b]
	\centering
	\includegraphics[width=\columnwidth]{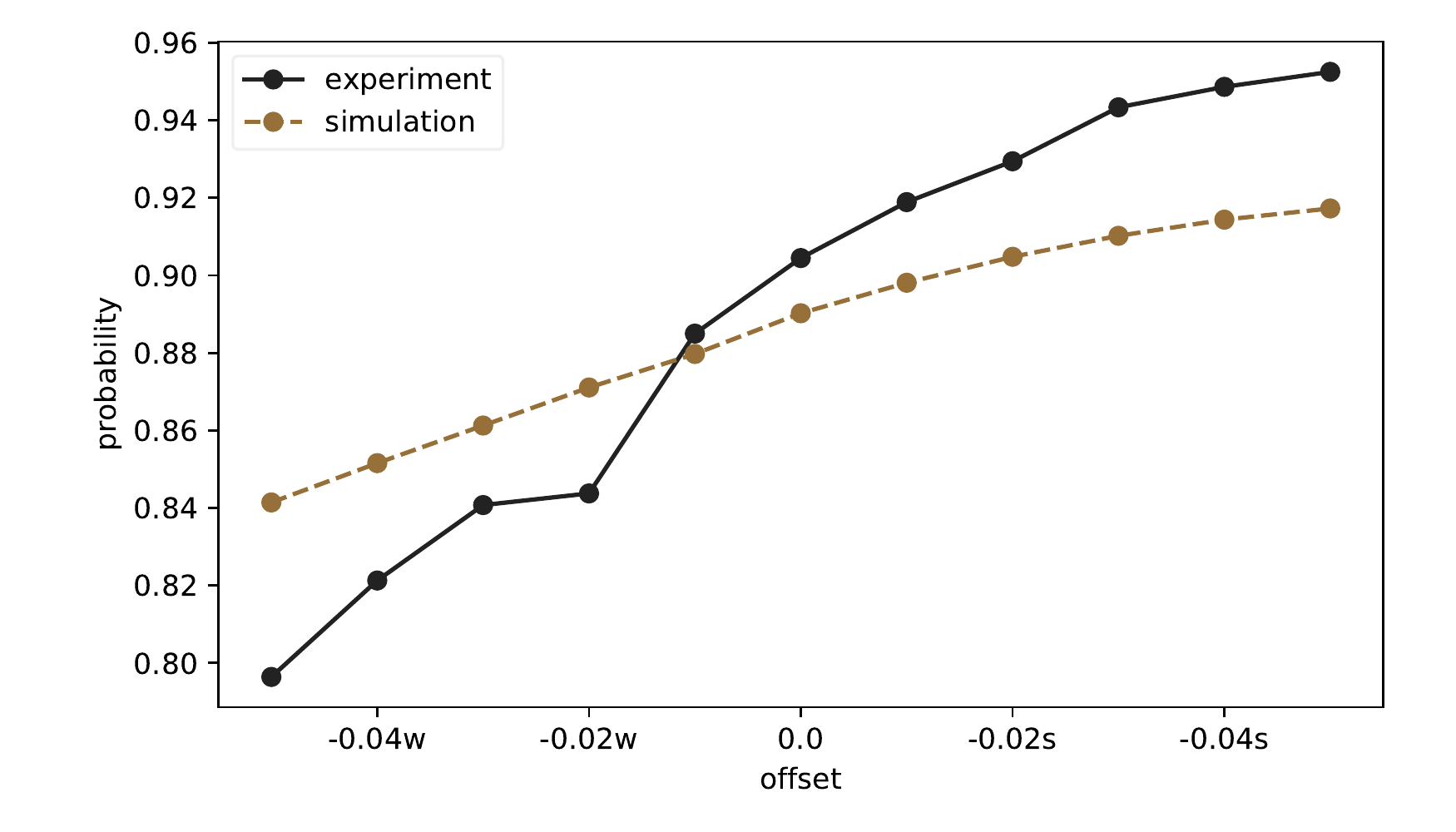}
	\caption{The probability of finding the MDS for $G(2)$ from D-Wave (black) and simulation (dashed yellow) at annealing times of 1 $\mu s$.}
	\label{fig:dwave1us}
\end{figure}

%----------------------------------------------------------------------------------------
\label{sec:results:simulation}
To understand the effects of annealing offsets, we simulate the annealing process for the $G(2)$ graph embedded in chimera topography.
We repeat the same anneal process with a shortend total annealing time of 1 $\mu s$, reducing the computational demands of the simulation, and count the number of correct ground state occurrences.
The resulting ground state probability as a function of offset measured over 100,000 observations is shown in Fig.~\ref{fig:dwave1us}.

To solve for time evolution dynamics of quantum annealing including thermal and the decoherence effects, we solve for the master equation in Lindblad form
\begin{align}
 \label{eq:sim:linblad-eq}
 \partial_t \rho (t) =  \frac{-i}{\hbar} [H(t) , \rho(t)] + \mathcal{L}(\rho(t), H(t))
\end{align}
where $\hbar$ is the reduced Plank's constant, $\rho (t)$ is the density matrix at time $t$, $H(t)$ is the time-dependent Hamiltonian and $\mathcal{L}$ is the Linblad operator implementing the decoherence models.
In this work, we consider two types of decoherence models: full-counting statistics~\cite{PhysRevE.90.022103,RevModPhys.81.1665} and single-qubit amplitude damping (local damping)~\cite{10.5555/1972505,preskill1998lecture}.
The full-counting statistics term models the global decoherence to all the qubits due to the classical reservoir.
The local damping term models the decoherence of each qubit independently.
Details of the master equation are given in Sec.~\ref{sec:methods:lindblad}.
Implementation of the annealing schedule and annealing offsets for the simulation are discussed in Sec.~\ref{sec:methods:annealing-schedule}.

\begin{figure}
	\centering
	\includegraphics[width=\columnwidth]{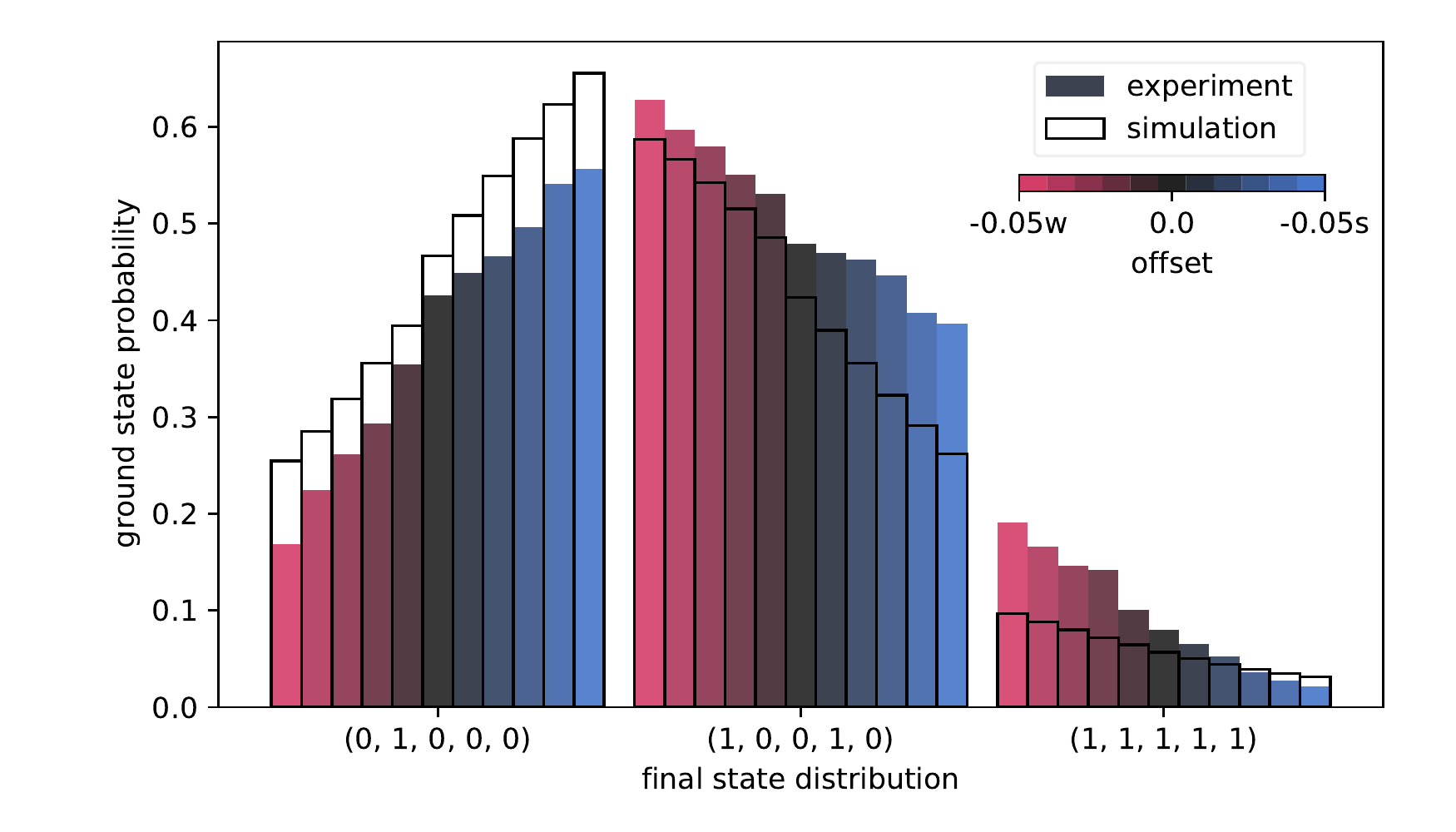}
	\caption{Final state distribution from D-Wave (solid bars) and simulation (black outline). The colors label the type of offset  The $(0, 1, 0, 0, 0)$ state is the first solution where vertex 1 is in the dominating set. The $(1, 0, 0, 1, 0)$ state is the second solution where vertex 0 is in the dominating set. The first excited state is the $(1, 1, 1, 1, 1)$ state where both vertices are in the dominating set.}
	\label{fig:final_state_distribution}
\end{figure}

The graph $G(2)$ has degenerate ground states depending on whether qubit 0 or 1 is chosen to be the MDS solution.
This degeneracy is reflected in the experimental result and provides a non-trivial benchmark for our simulation.
Fig.~\ref{fig:final_state_distribution} show the final state distribution of the three lowest-lying state.
The states $(0, 1, 0, 0, 0)$ and $(1, 0, 0, 1, 0)$ are the two degenerate ground states of the embedded Hamiltonian, while $(1, 1, 1, 1, 1)$ is the first excited state which yields an incorrect solution\footnote{According to Eq.~\eqref{eq:bit-vector-mapping}, the first two vector components represent the nodes of the graph while the following components represent the slack variables after embedding.}.
All other states receive negligible probability at the end of annealing.
The simulation (black) captures the main features of the experimental result:
\begin{enumerate}
    \item Significant probability of populating both ground states (rather than populating only one)
    \item Asymmetry in ground state population due to offsets and spanning approximately the correct range
    \item Population of first excited state with systematically lower probability when the strong field is delayed
\end{enumerate}
The asymmetry in the ground state distribution at non-zero offset results from annealing offsets lifting the ground state degeneracy.
The non-degenerate ground state switches between the two states within the broken symmetry, depending on which qubit group is delayed. At zero offset, a slight asymmetry exists in the simulation because the Hamiltonian is only degenerate at the last moment, while the D-Wave is also subject to sampling bias~\cite{2016PhRvA..93e2320K, 2017PhRvL.118g0502M}.

This result can be obtained by tuning three free parameters: the simulation temperature to the order of 10 milliKelvin, and the two coefficients of the two decoherence models at the order of 1 to 10 $ns$. These values are consistent with the reported D-Wave operating temperature~\cite{dwave_temp} and coherence times for flux qubits~\cite{2003Sci...299.1869C}.
Additional insights of the simulation are given in Sec.~\ref{sec:discussion:time_evolution}.

The resulting probability of recovering the correct solution as a function of annealing offset is given in Fig.~\ref{fig:dwave1us}.
We confirm that the simulation's offset-scaling follows the experiment's scaling, which suggests that the improvements are related to quantum mechanics.
An additional study where an extended annealing schedule is employed in the simulation, which removes systematic errors introduced by annealing offsets and effects of local damping are presented in Sec.~\ref{sec:discussion:idealqa}, and further supports this observation.

\section{Discussion}
\subsection{Dynamics of Time Evolution}
\label{sec:discussion:time_evolution}

\begin{figure}
	\centering
	\includegraphics[width=\columnwidth]{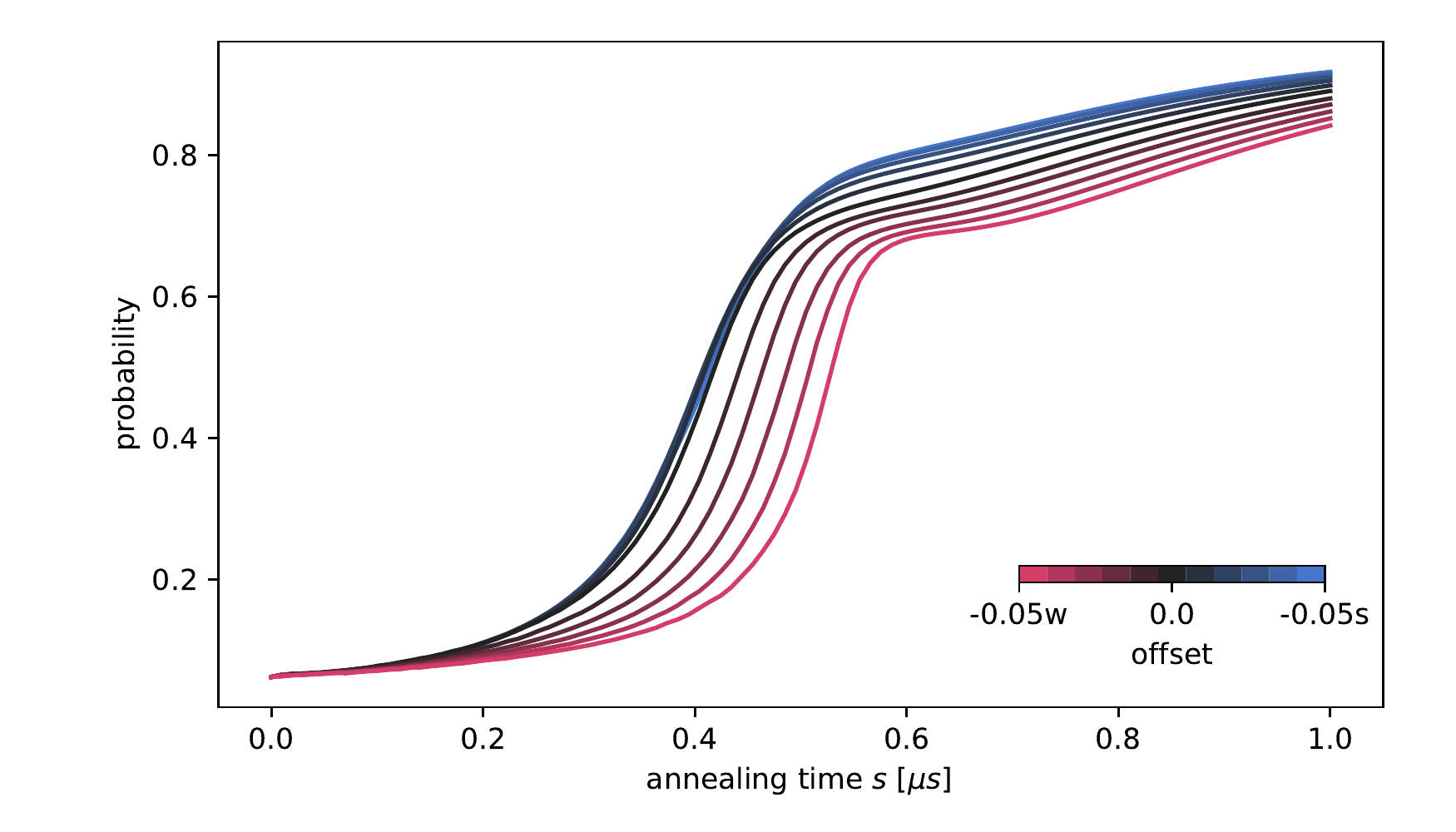}
	\caption{Time-dependent probability for resolving the ground state for simulation results presented in Sec.~\ref{sec:results:simulation}. Different offsets are labeled in the same way as Fig.~\ref{fig:baseline}.}
	\label{fig:td_prob}
\end{figure}

The time-dependent overlap with the exact Ising ground state is shown in Fig.~\ref{fig:td_prob} from applying the simulator to $G(2)$. We observe for all cases that the system undergoes what is analogous to a magnetic phase transition around $s\sim 0.4$.
After the phase transition, we can confirm that the system collapses to effectively a classical state in the sense that the density matrix becomes a diagonal matrix.
The steady increase in probability after the phase transition is a sensitive balance between the competing effects between full-counting statistics and local damping.
In our example, the full-counting statistics decoherance rate is tuned to be slightly stronger compared to the local dampening decoherance rate, effectively resulting in a final thermal annealing process after the phase transition.
If local damping were relatively larger, then the probability after the phase transition will slowly decrease as the system decoheres into its local ground state.
While we believe both effects are essential to simulate D-Wave, due to the competing effects of both decoherence models, we emphasize that a fully quantitative comparison of both decoherence models cannot be made just considering the $G(2)$ graph.
However, we emphasize that the simulation suggests that the ground state is recovered predominantly due to the quantum phase transition happing around $s\sim 0.4$.

Finally, we comment that effects of full-counting statistics are required for the simulation to obtain the final state distribution shown in Fig.~\ref{fig:final_state_distribution}.
Because of the smallness of the $G(2)$ problem (and the utilized total annealing time of 1$\mu s$), the annealing offsets lift the ground state degeneracy in a manor that diabatic transitions do not occur--setting a discrete, fixed groundstate probability for given offsets.
Thus, if dynamical thermalization effects were absent, the simulation would populate predominantly one of the two unique ground states depending on the offsets.
Once dynamical thermalization effects are present, the final state distribution  continously depends on offsets and becomes a very sensitive observable to tune the simulation temperature, which, if properly tuned, agrees well with the experimental operating temperature.

\subsection{Idealized Quantum Annealing}
\label{sec:discussion:idealqa}

\begin{figure*}
    \centering
	\begin{tabular}{p{0.5\textwidth}p{0.5\textwidth}}
	\includegraphics[width=0.5\textwidth]{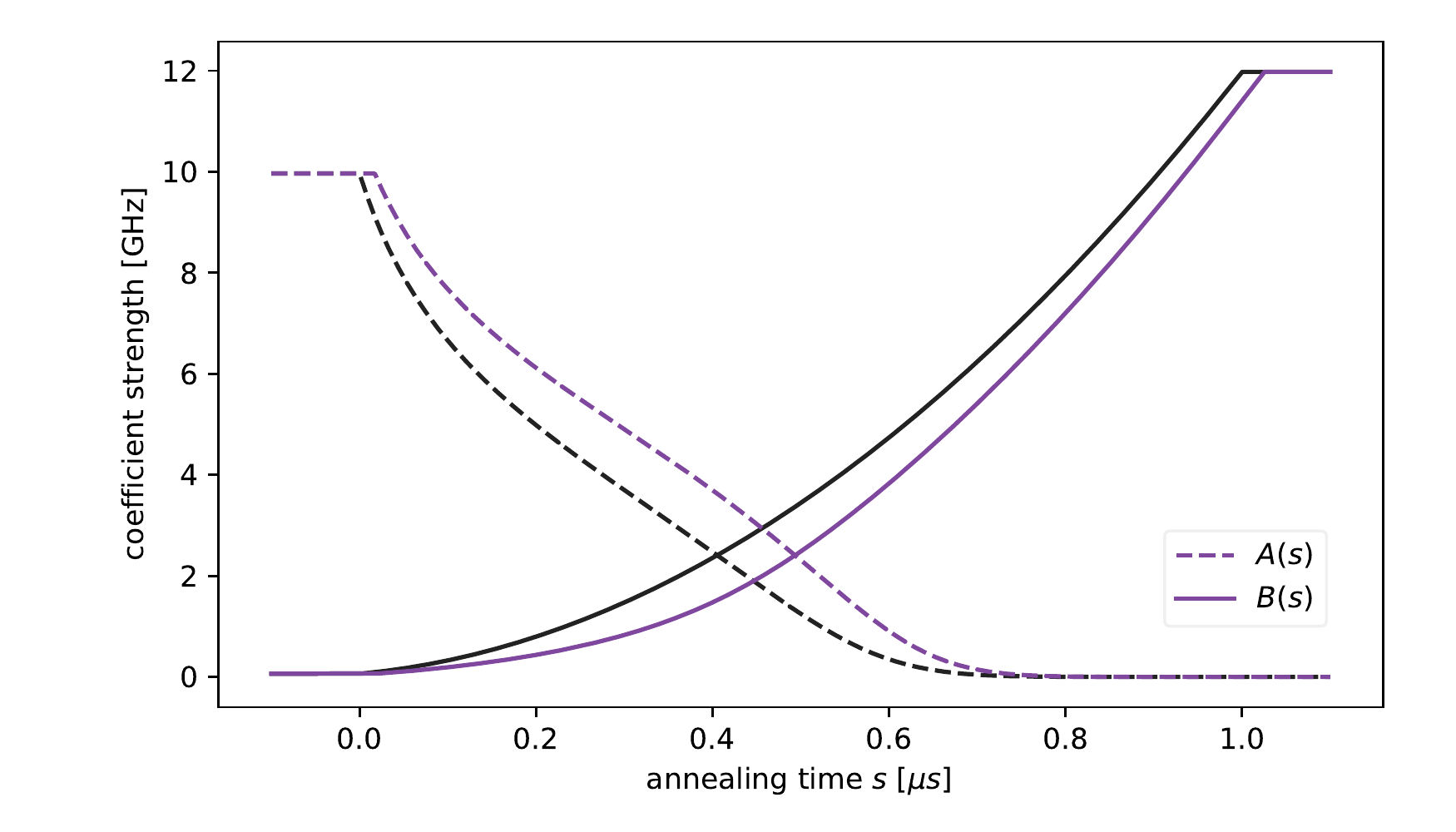}
    &
	\includegraphics[width=0.5\textwidth]{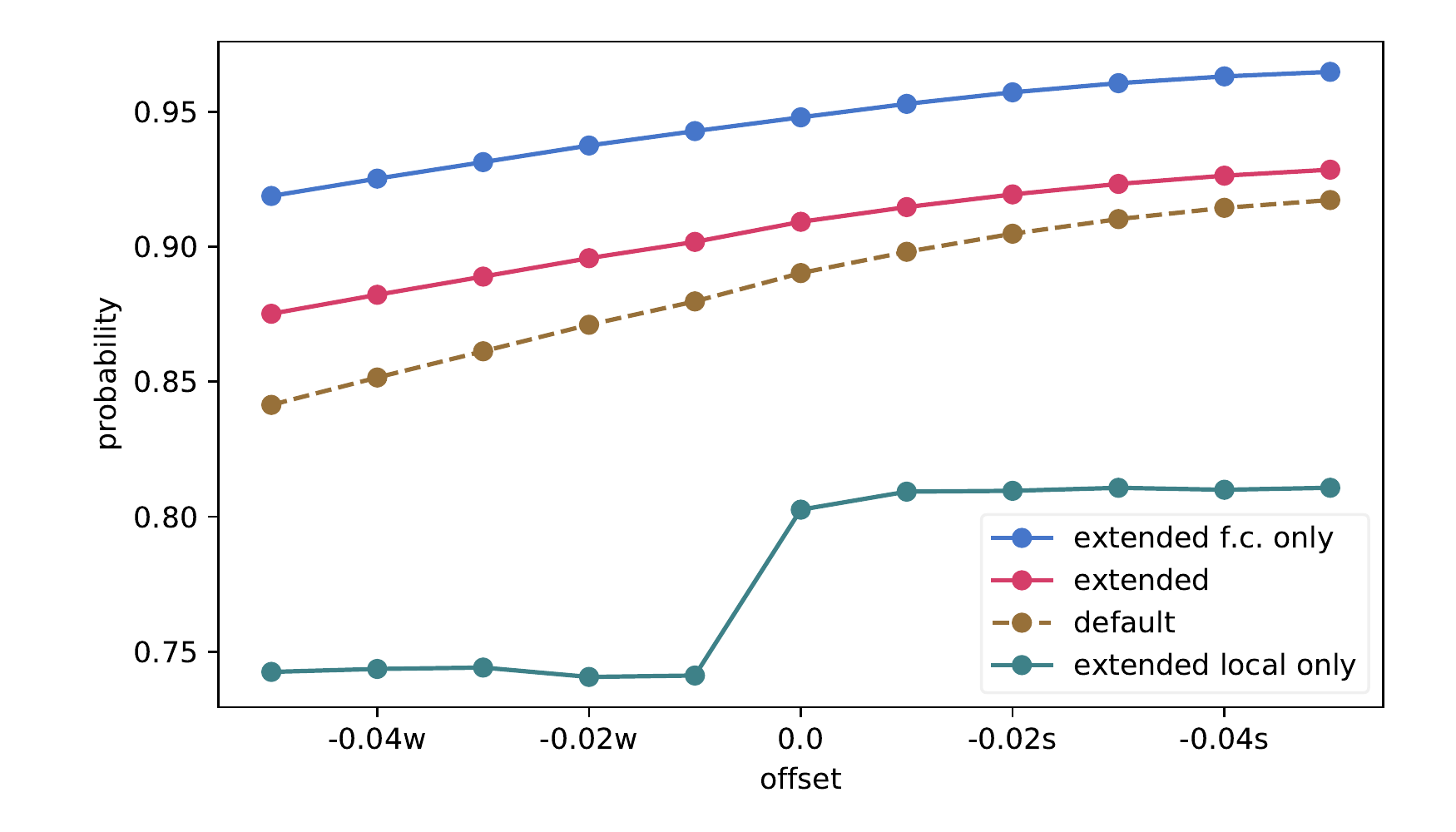}\\
	\centering \textbf{a} & \centering \textbf{b}
	\end{tabular}
	\centering
	\caption{Extended annealing schedule and simulation result. \textbf{a)} The annealing time is increased by 10\% at both the start and end, and the initial values of $A(s)$ and $B(s)$ at non-zero offset are also not extrapolated in comparison to Fig.~\ref{fig:anneal_schedule}. Therefore, the initial and final Hamiltonian are no longer suffer systematic error at non-zero offset. \textbf{b)} The default (dashed yellow) result is one presented in Fig.~\ref{fig:dwave1us}. Results from extended schedules are presented shown with solid lines with full-counting statistics and amplitude damping (red), only full-counting-statistics (blue), and only amplitude damping (green).
    }
	\label{fig:anneal_schedule_ext}
\end{figure*}

In the simulation, we also reserve the ability to remove systematic errors associated with finite annealing schedules by extending them, as shown by the red data points in Fig.~\ref{fig:anneal_schedule_ext}. With the extended schedule, all qubits start and end with the same values of $A$ and $B$ and faithfully preserves the initial and final Hamiltonians. The dependence on offset for $G(2)$ (Fig.~\ref{fig:dwave1us}) remains the same under the extended anneal schedule (Fig.~\ref{fig:anneal_schedule_ext}), confirming that the behavior is not a systematic artifact.

The quantum system can be further idealized by including either solely the full-counting statistics model (blue) or solely the local decoherence model (green)\footnote{For problems without diabatic transitions, like the MDS problem for a $G(2)$ graph and an anneal time of $1\mu s$, removing all decoherence models results in ground state probabilities equal to unity.}. We observe in both cases that the overall scaling follows the story of the MBL hypothesis. Perhaps more importantly, we observe that a simulation without local decoherence, where the relaxation is dependent on precisely the instantaneous value of $|h_i|$, exhibits the same scaling as the experiment. This result suggests that our strategy for setting annealing offsets is improving the algorithm beyond the simple explanation of qubits freezing due to single-particle amplitude damping. The simulation with only amplitude damping (green) does not fully capture the results of the experiment. We observe hints of a phase transition depending on the offset (which may be considered some measure of disorder) when dynamical thermalization effects are removed. This observation is a strong evidence for the inclusion of the full-counting statistics model.

\subsection{Final Remarks}
\label{sec:results:final}
We want to emphasize that while the annealing offsets are motivated by the MBL hypothesis, and the results also follow those expectations, we do not have definitive proof that MBL plays a crucial role.
Observations of MBL inevitably require the study of finite-size scaling~\cite{2015PhRvB..91h1103L}, and our current simulation, while being extremely thorough and explicit, is exponentially slow to evaluate, making evaluations of even $G(3)$ unfeasible.
However, the intersection of time-dependent quantum mechanics and emergent phenomena~\cite{Goldenfeld:1992qy} is an exciting direction that is pertinent to adiabatic quantum computing.

%========================================================================================
\section{METHODS}
\label{sec:methods}
%========================================================================================
%----------------------------------------------------------------------------------------
\subsection{Additional Mappings to QUBO}
\label{sec:methods:ILP-to-QUBO}
%----------------------------------------------------------------------------------------
%----------------------------------------------------------------------------------------
\subsubsection{Minimum Dominating Set QUBO}
\label{sec:methods:mds-qubo}
%----------------------------------------------------------------------------------------

The solution to the MDS problem can be expressed as an integer optimization problem given by,
\begin{align}
 f(\vec x) = & \min\left(\sum_{v \in V} x_v\right),                    \\
\end{align}
subject to
\begin{align}
 & x_v + \sum_{j \in \mathcal{N}(v)} x_j \geq 1, \\
 & x_v \in \{0, 1\}
\end{align}
where $\mathcal{N}(v)$ represents the set of all direct neighbors of vertex $v$ and the dimension of the dependent variable $x$ is the number of vertices $n_V \equiv \overline{\overline{V}}$.
The problem minimizes the number of vertices in $D$ with a binary variable $\vec x = \vec \Psi_x$ encoded by single qubit, subject to the constraint that at least one vertex in $\mathcal{N}(\nu)$ is in $D$.
For each vertex in $V$ we introduce slack variables
\begin{equation}
    \vec s \in \left\{ \mathbb{Z}^{n_V} \, \middle| \, 0 \leq s_{v} \leq |\mathcal{N}(v)| \quad \forall v\in V \right\} \, ,
\end{equation}
which is related to the qubit vector $\vec \Psi_s$ by $\vec s = T_s \vec \Psi_s$.
The inequality constraint is encoded by
\begin{align}
 f(\vec \Psi_x)
 =
 & \min\left(\vec 1 \cdot \vec \Psi_x \right) \, ,
\end{align}
subject to
\begin{align}
 &
 (\mathbbm{1} + J)\vec \Psi_x - T_s \Psi_s  - \vec 1 = 0\,,
 \\
 &
 (\vec \Psi_s)_\nu \in \{ 0, 1\}
\end{align}
where the nearest-neighbor sum is expressed by the adjacency matrix $J$ (zero diagonal and symmetric for non-directional graphs).
The algorithm uses
\begin{equation}
    N_q = \overline{\overline{V}} + \sum_{v \in V} \log_2 \mathcal{N}(v)
\end{equation}
qubits to encode the vertices and slack variables before embedding.
Therefore, the (logical) binary vector $\vec \Psi$ at worst scales with $n_V \log_2 n_V$ qubits for fully connected graphs.

The target QUBO in the notation of Eq.~\eqref{eq:matrix_form} reads
 {\small
  \begin{align}
   Q_{xx} & = \mathbbm{1} + p\left[J^T J + J^T + J - \mathrm{diag}\left(J^T \vec 1 + \vec 1^T J\right) - \mathbbm{1} \right] \,, \\
   Q_{sx} & = - p(\mathbbm{1}+J)^TT_s\,,                                                                     \\
   Q_{ss} & = p\left[{T_s}^T T_s + \mathrm{diag}\left(T_s^T \vec 1 + \vec 1^T  T_s\right)\right]\,,                                  \\
   C      & =  p \overline{\overline{V}}\,.
  \end{align}}

%----------------------------------------------------------------------------------------
\subsubsection{Integer Quadratic Optimization}
\label{sec:methods:ilp:quadratic}
%----------------------------------------------------------------------------------------

Because quantum annealing is capable of solving quadratic problems, we extend the proposed algorithm to solve integer quadratic optimization problems as well such that
\begin{align}
 \vec x_0 = \mathrm{arg}\min\limits_{x}\left(\sum_{ij} x_i d_{ij} x_j + \sum_i c_i x_i\right)
\end{align}
without the introduction of auxiliary qubits.

In this case the $Q_{xx}$ component of the QUBO, Eq.~\eqref{eq:qubo:components} obtains a new term
\begin{equation}
    Q_{xx} \to Q_{xx} + T_x^T d T_x \, .
\end{equation}

%----------------------------------------------------------------------------------------
\subsection{ILP on the D-Wave}
\label{sec:methods:ILP-on-D-Wave}
%----------------------------------------------------------------------------------------

%----------------------------------------------------------------------------------------
\subsubsection{Comment on ILP QUBO Penalty Term}
\label{sec:methods:ilp-qubo-comments}
%----------------------------------------------------------------------------------------

The minimal energy solution to Eq.~\eqref{eq:initial-ip-def} and Eq.~\eqref{eq:ilp-slack-bit-energy} are exactly the same if the penalty term is greater or equal to the energy gap of the first excited solution: $p \geq E_1 - E_0$.
Thus some knowledge of the problem is required.
In principle, it is possible to set the penalty term arbitrarily large, at the cost of problem resolution: large values for $p$ increase the highest available energy of the system by multiples of $p$.
After normalization of the QUBO, this corresponds to decreasing the energy gap between the ground state and the first excited state.
Thus, if solvers have finite precision, one must estimate reasonable values for $p$: for large $p$ more solutions fulfill the constraints, while for small $p$, more solutions minimize the objective function.

%----------------------------------------------------------------------------------------
\subsubsection{Minor Embedding}
\label{sec:methods:minor_embedding}
%----------------------------------------------------------------------------------------
Our proposed offset strategy is motivated by the structure of the Hamiltonian being evaluated by the annealer.
As a result, details of the embedding are important. We obtain an embedding for $G(n)$ with the \texttt{embed\_qubo} function provided by the D-Wave Ocean Python package~\cite{dwave_oceans} under the \texttt{dwave.embedding} module~\cite{2008arXiv0804.4884C}.
The same embedding is used for all D-Wave solves of the same graph (independent of offset), and consequently the simulation solves the resulting embedded Hamiltonian for $G(2)$.
Additionally, solving the same graph as a function of offset on the exact same qubits removes (or at least keeps consistent) the systematic effects due to solving a problem on different physical qubits.
We note that comparisons between different graphs in Fig.~\ref{fig:baseline} are subject to this uncontrolled systematic.

After embedding the QUBO for $G(2)$ requires 5 qubits (an increase from 4), where by construction, qubits 0 and 3 form the qubit chain.
We confirm through brute force evaluation of the eigenvalue decomposition of the 5 qubit Hamiltonian, that the embedding provided by D-Wave solves the expected ILP problem for $G(2)$, with degenerate ground states at $(0, 1, 0, 0, 0)$ and $(1, 0, 0, 1, 0)$ corresponding to whether vertex 0 or 1 is chosen for the MDS solution, and $(1, 1, 1, 1, 1)$ as the first (non-degenerate) excited state where both vertex 1 and 0 are in the set yielding a valid dominating set but not the MDS.

The resulting Ising Hamiltonian has external field equal to $h = (2.75, 1.5, -1.0, -1.25, -1.0)$.
Following the offset strategy described in Sec.~\ref{sec:results:qa}, qubit(s) 0 (1, 2, 3, 4) are placed in the set with relatively stronger (weaker) final external fields.
This imbalance in the two groups may perhaps explain the reason why effects of delaying the weaker fields are more pronounced, since delaying the strong fields only differs from the baseline by a single qubit.

%----------------------------------------------------------------------------------------
\subsection{Simulation of a Quantum Annealer}
\label{sec:methods:simulation}
%----------------------------------------------------------------------------------------

%----------------------------------------------------------------------------------------
\subsubsection{The Lindblad Equation}
\label{sec:methods:lindblad}
%----------------------------------------------------------------------------------------

To solve for time evolution dynamics of quantum annealing including thermal and the decoherence effects, we evaluate the master equation Eq.~\eqref{eq:sim:linblad-eq} in Lindblad form.
The explicit time-dependence of the Hamiltonian is given by
\begin{align}
 \label{eq:annealH}
 H(t)  = & - \sum_i  A_i(t)\sigma^x_i +\sum_i B_{i}(t) h_i \sigma^z_i \notag \\
 & + \sum_{i>j} \sqrt{B_{i}(t)B_{j}(t)} J_{ij} \sigma^z_i \sigma^z_j  ,
\end{align}
where $A_i(t)$ and $B_{i}(t)$ are site-dependent annealing schedule functions.
The site dependency takes into account of the annealing offsets.
The simulation takes two decoherance models into account.
Full-counting statistics and local decoherence.

For full-counting statistics, the Lindblad operator is
\begin{align}
\mathcal{L}_{\textrm{fc}}(\rho (t), H(t)) = & \Gamma_{\textrm{fc}} \sum_{j, i<j} \left[  \left(2 S_{ij} \rho(t) S_{ij}^\dagger - \{ S^\dagger_{ij} S_{ij}, \rho(t) \}\right) \notag \right.
\\
& \left. + e^{-\beta \Delta E_{ij}} \left(2 S_{ij}^\dagger \rho(t) S_{ij} - \{ S_{ij} S^\dagger_{ij}, \rho(t) \} \right)\right]
\end{align}
where $\{, \}$ denotes the anti-commutator, $(ij)$ is the index for the inter-level spacing $\Delta E_{ij}=E_j-E_i>0$,  $S_{ij}=|E_i \rangle \langle E_j|$ denotes the many-body lowering operator and $\Gamma_{fc} = 1/T_{fc}$ is the full-counting decoherence rate for coherence time $T_{fc}$.
That is, due to the interaction with the classical thermal bath, there is a probability that the system hops from each higher-energy many-body state to a lower-energy many-body state.
The probability of the inverse process is given by a Boltzmann factor.

To model the local decoherence of each qubits in the non-interacting limit, we also consider the amplitude damping for non-interacting qubits.
For the local decoherance model, the Lindblad operator is
\begin{align}
\mathcal{L}_{\textrm{loc}}(\rho (t), H(t)) = &  \Gamma_{\textrm{loc}} \sum_j \left(2L_j \rho(t) L_j^\dagger - \{ L^\dagger_j L_j, \rho(t) \} \right)\notag \\
& + e^{- 2 \beta |h_j| }  \left(2L_j^\dagger \rho(t) L_j - \{ L_j L_j^\dagger, \rho(t) \} \right) ,
\end{align}
where $j$ is the index for qubit.
$L_j= \sigma^{+}_j=|1\rangle \langle 0|$ for $h_j>0$ and $L_j= \sigma^{-}_j=|0\rangle \langle 1|$ for $h_j<0$.
That is, each qubit is damped toward its local ground state if we ignore all the interactions.

The initial condition is the Gibbs canonical ensemble
\begin{equation}
\rho (0) =  \frac{e^{-\beta H(0)}}{\mbox{Tr}\left(e^{-\beta H(0)}\right)} ,
\end{equation}
where $\beta$ is the inverse temperature.
The probability to get the ground state at measurement is
\begin{align}
P =  \mbox{Tr} \left( \rho (t) \pi_{\mbox{gnd}} \right)  ,
\end{align}
where the projection operator onto the degenerated ground states subspace is defined as $\pi_{\mbox{gnd}}=\sum_{i\in G} |\mbox{gnd}_i\rangle \langle \mbox{gnd}_i| $.
Here $\{ | \mbox{gnd}_i \rangle | i \in G \}$ forms an orthonormal basis for the degenerated ground states subspace, i.e., $\langle \mbox{gnd}_j | \mbox{gnd}_i \rangle = \delta_{ij}$.

For the simulation results presented, we set the temperature to 22.5 milliKelvin.
The coherence time for local decoherence is set to 15$ns$, and 1$ns$ for full-counting statistics.

%----------------------------------------------------------------------------------------
\subsubsection{Annealing Schedule and Offsets}
\label{sec:methods:annealing-schedule}
\begin{figure}[htb]
 \centering
  \includegraphics[width=\columnwidth]{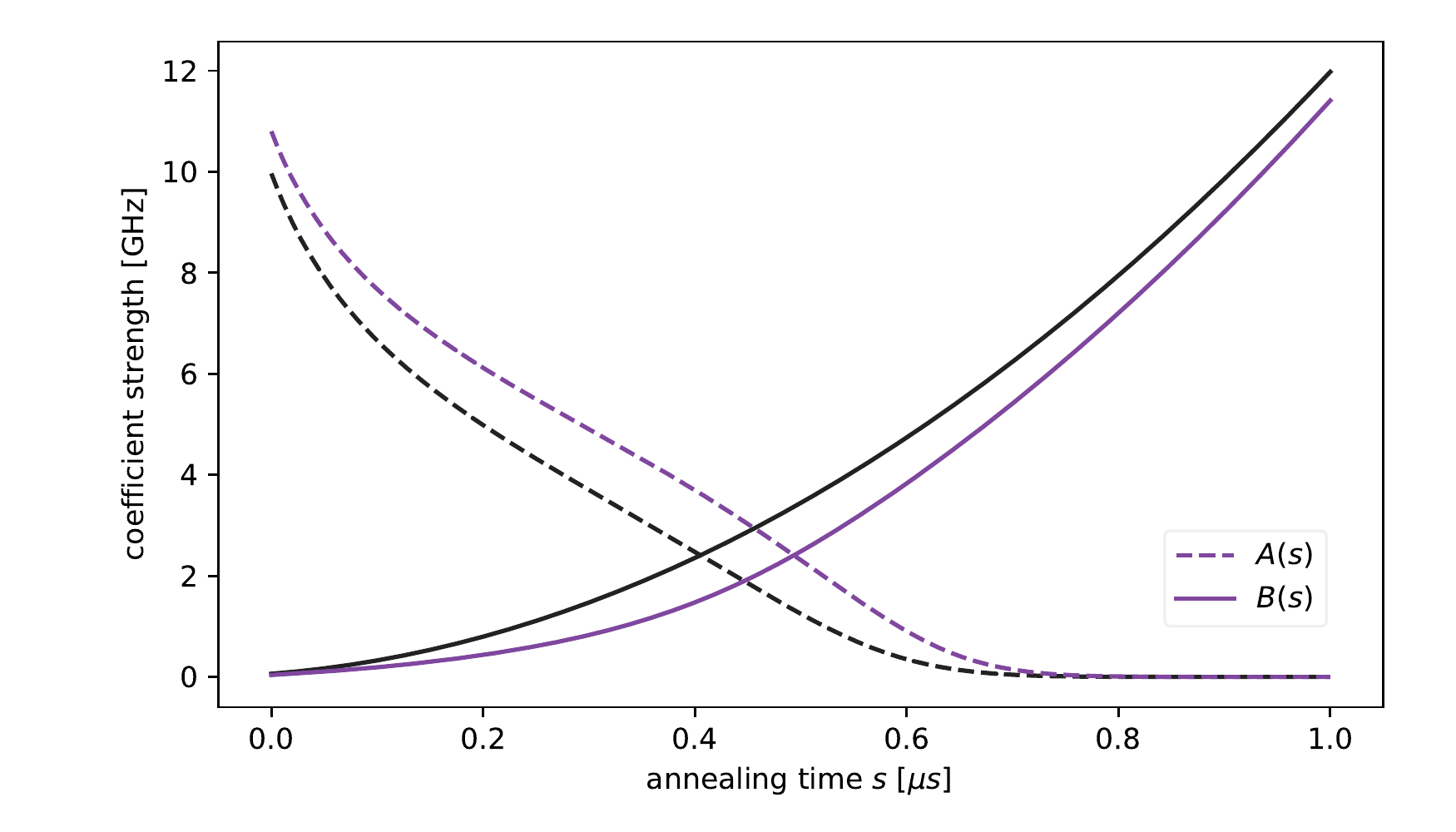}
  \caption{
  Anneal schedules for amplitudes of initial Hamiltonian (dashed) and final Hamiltonian (solid).
  Offset 0.0 (black) and -0.05 (purple) are displayed.
  Intermediate values lie between and are suppressed for clarity.
 }
 \label{fig:anneal_schedule}
\end{figure}

%----------------------------------------------------------------------------------------
In this section, we discuss the details of the annealing schedule with respect to the dimensionless normalized time $s$.
On D-Wave solvers, annealing offsets effectively advance or delay the annealing schedule of individual qubits (see E.~\eqref{eq:annealH}).
In Fig.~\ref{fig:anneal_schedule}, the default D-Wave annealing schedule is shown in black, in addition to the effects of applying negative offsets (effective time delay) to $A(s)$ and $B(s)$ in blue.
Further documentation is provided by D-Wave in Ref.~\cite{dwave_as, dwave_as_docu}.

The coefficients $A(s)$ and $B(s)$ follow the underlying control variable $c(s)$, which is designed to grow the persistent current $I_p(s)$ linearly in time.
The effective time delay is implemented by introducing an offset as $c(s) + \delta$.
Because annealing schedules are finite, systematic errors are introduced because the final values of $A(s)$ and $B(s)$ will differ for qubits with different offsets.
Additionally, the values of the coefficients are unknown outside of $s\in [0, 1]$.
We only employ negative offsets such that this unkonwn coefficient range, approximated by a linear extrapolation, only enters at the beginning of the annealing process.

We verify that our implementation of annealing offsets on the simulator is consistent with D-Wave by solving the following three qubit Hamiltonian
\begin{align}
	\label{eq:offset_test_hamiltonian}
	H^{\textrm{problem}} =
	\begin{pmatrix}
		-0.25 & 1 & 0 \\
		0 & -0.25 & 0 \\
		0 & 0 & -0.25
	\end{pmatrix}
\end{align}
which has a doubly-degenerate ground state of $(0, 1, 1)$ and $(1, 0, 1)$. An annealing offset is then applied to either qubit 0 or 1, and thus breaks the ground state degeneracy of the system. Because of the systematic error introduced when assigning an offset to a qubit, the final Hamiltonian given by Eq.~(\ref{eq:annealH}) will have small deviations from the input problem Hamiltonian. For example, we expect a negative offset to qubit 0 to Eq.~(\ref{eq:offset_test_hamiltonian}) will yield $(1, 0, 1)$ as the unique ground state given Eq.~(\ref{eq:annealH}).

We confirm that with different combinations of annealing offsets, the degeneracy is lifted on the D-Wave results as expected by solving for the modified problem Hamiltonian spectrum, as well as dialing in annealing offset in the simulation of this three qubit test case.

%----------------------------------------------------------------------------------------
\subsubsection{Pure Transverse Field Simulation}
\label{sec:methods:simulation_details}
\begin{figure}
	\centering
	\includegraphics[width=\columnwidth]{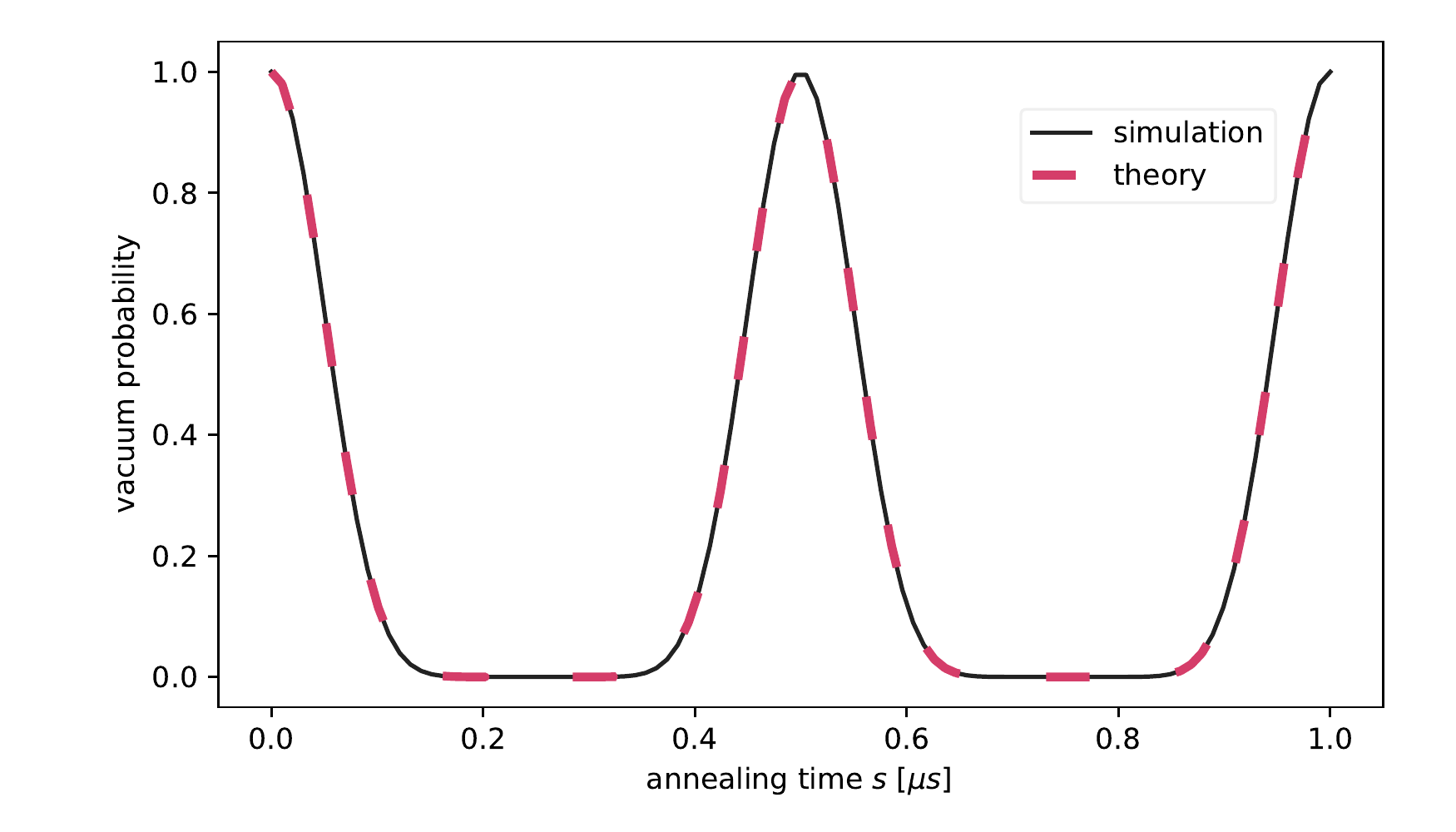}
	\caption{Time-dependent vacuum probability of 5 qubits system under pure transverse field.}
	\label{figcheck}
\end{figure}

%----------------------------------------------------------------------------------------
To check the correctness of simulator and unit conversion, we tested a simple annealing schedule for pure transverse field, i.e. $A(s)=A(0)$ and $B(s)=0$.
The initial state is pure zero state $\rho=|00...0\rangle \langle 00...0|$.
In this case the analytical solution can be obtained for the wave function oscillation.
The time-dependent probability is $P_{0}=\cos^{2n}(s)=\cos^{2n}(t/T)$ where $n$ is the number of qubits and $T$ is the annealing time.
For annealing time $T=1/A(0)$, we expect to see perfect one-period oscillation. The energy spectrum for this system can be analytically obtained, so we also checked the Boltzmann distribution in the initial density matrix construction. The oscillation is depicted in Fig.~\ref{figcheck}, where the simulation matches the expected theoretical behavior.

%========================================================================================
\section{DATA AVAILABILITY}
\label{sec:open-source}
%========================================================================================
We provide access to all the software utilized and data presented in this publication in the associated GitHub repository~\cite{github:cchang5/quantum_linear_programming}.
The utilized software is contained in two installable Python modules:
\begin{itemize}[leftmargin=*]
    \itemsep0em
    \item[] \texttt{qlp} for mapping MDS problems to QUBOs and performing simulations,
    \item[] \texttt{qlpdb} for interfacing and documenting simulation and experiment data through EspressoDB~\cite{Chang:2019khk}.
\end{itemize}
The data itself is stored in a SQLite file and associated repository subdirectories.
We provide further documentation in the repository \texttt{readme} file.

%========================================================================================
\section{ACKNOWLEDGEMENTS}
%========================================================================================

We thank Long Gang Pang for initial discussions.
We thank David Johnson, Vlad Papish and the rest of D-Wave Support for answering detailed questions about the D-Wave annealer.

This material is based upon work supported by the U.S. Department of Energy, Office of Science, Office of Nuclear Physics, Quantum Horizons: QIS Research and Innovation for Nuclear Science under Award Number FWP-NQISCCAWL.
Lawrence Berkeley National Laboratory (LBNL) is operated by The Regents of the University of California (UC) for the U.S. Department of Energy (DOE) under Federal Prime Agreement DE-AC02-05CH11231.
C.K. gratefully acknowledges funding through the Alexander von Humboldt Foundation through a Feodor Lynen Research Fellowship.
This work used resources of the Oak Ridge Leadership Computing Facility, which is a DOE Office of Science User Facility supported under Contract DE-AC05-00OR22725.
This manuscript has been authored by UT-Battelle, LLC under Contract No. DE-AC05-00OR22725 with the U.S. Department of Energy.
The United States Government retains and the publisher, by accepting the article for publication, acknowledges that the United States Government retains a non-exclusive, paid-up, irrevocable, world-wide license to publish or reproduce the published form of this manuscript, or allow others to do so, for United States Government purposes. The Department of Energy will provide public access to these results of federally sponsored research in accordance with the DOE Public Access Plan. (http://energy.gov/downloads/doe-public-access-plan).

%========================================================================================
\section{AUTHOR CONTRIBUTIONS}
%========================================================================================

Initial idea was proposed by Chang.
Ostrowski guided the ILP problem selection.
All authors contributed to the design of test cases.
Calculations and code development for the D-Wave were performed by Chang and K\"orber.
Code for numerical simulations was developed by Chen and K\"orber, and cross-checked by Chang.
Simulations were performed by Chang and Chen.
Chang, Chen, K\"orber, and Humble interpreted the results.
All authors contributed to writing and editing of the final manuscript.

%========================================================================================
\section{ADDITIONAL INFORMATION}
%========================================================================================

\textbf{Competing Interests:} The authors declare no competing interests.

\bibliographystyle{apsrev4-1}
\bibliography{main.bib}

\begin{thebibliography}{35}%
\makeatletter
\providecommand \@ifxundefined [1]{%
 \@ifx{#1\undefined}
}%
\providecommand \@ifnum [1]{%
 \ifnum #1\expandafter \@firstoftwo
 \else \expandafter \@secondoftwo
 \fi
}%
\providecommand \@ifx [1]{%
 \ifx #1\expandafter \@firstoftwo
 \else \expandafter \@secondoftwo
 \fi
}%
\providecommand \natexlab [1]{#1}%
\providecommand \enquote  [1]{``#1''}%
\providecommand \bibnamefont  [1]{#1}%
\providecommand \bibfnamefont [1]{#1}%
\providecommand \citenamefont [1]{#1}%
\providecommand \href@noop [0]{\@secondoftwo}%
\providecommand \href [0]{\begingroup \@sanitize@url \@href}%
\providecommand \@href[1]{\@@startlink{#1}\@@href}%
\providecommand \@@href[1]{\endgroup#1\@@endlink}%
\providecommand \@sanitize@url [0]{\catcode `\\12\catcode `\$12\catcode
  `\&12\catcode `\#12\catcode `\^12\catcode `\_12\catcode `\%12\relax}%
\providecommand \@@startlink[1]{}%
\providecommand \@@endlink[0]{}%
\providecommand \url  [0]{\begingroup\@sanitize@url \@url }%
\providecommand \@url [1]{\endgroup\@href {#1}{\urlprefix }}%
\providecommand \urlprefix  [0]{URL }%
\providecommand \Eprint [0]{\href }%
\providecommand \doibase [0]{http://dx.doi.org/}%
\providecommand \selectlanguage [0]{\@gobble}%
\providecommand \bibinfo  [0]{\@secondoftwo}%
\providecommand \bibfield  [0]{\@secondoftwo}%
\providecommand \translation [1]{[#1]}%
\providecommand \BibitemOpen [0]{}%
\providecommand \bibitemStop [0]{}%
\providecommand \bibitemNoStop [0]{.\EOS\space}%
\providecommand \EOS [0]{\spacefactor3000\relax}%
\providecommand \BibitemShut  [1]{\csname bibitem#1\endcsname}%
\let\auto@bib@innerbib\@empty
%</preamble>
\bibitem [{\citenamefont {Glover}(1986)}]{GLOVER1986533}%
  \BibitemOpen
  \bibfield  {author} {\bibinfo {author} {\bibfnamefont {F.}~\bibnamefont
  {Glover}},\ }\href {\doibase https://doi.org/10.1016/0305-0548(86)90048-1}
  {\bibfield  {journal} {\bibinfo  {journal} {Computers \& Operations
  Research}\ }\textbf {\bibinfo {volume} {13}},\ \bibinfo {pages} {533 }
  (\bibinfo {year} {1986})},\ \bibinfo {note} {applications of Integer
  Programming}\BibitemShut {NoStop}%
\bibitem [{\citenamefont {Glover}(1989)}]{doi:10.1287/ijoc.1.3.190}%
  \BibitemOpen
  \bibfield  {author} {\bibinfo {author} {\bibfnamefont {F.}~\bibnamefont
  {Glover}},\ }\href {\doibase 10.1287/ijoc.1.3.190} {\bibfield  {journal}
  {\bibinfo  {journal} {ORSA Journal on Computing}\ }\textbf {\bibinfo {volume}
  {1}},\ \bibinfo {pages} {190} (\bibinfo {year} {1989})},\ \Eprint
  {http://arxiv.org/abs/https://doi.org/10.1287/ijoc.1.3.190}
  {https://doi.org/10.1287/ijoc.1.3.190} \BibitemShut {NoStop}%
\bibitem [{\citenamefont {Glover}(1990)}]{doi:10.1287/ijoc.2.1.4}%
  \BibitemOpen
  \bibfield  {author} {\bibinfo {author} {\bibfnamefont {F.}~\bibnamefont
  {Glover}},\ }\href {\doibase 10.1287/ijoc.2.1.4} {\bibfield  {journal}
  {\bibinfo  {journal} {ORSA Journal on Computing}\ }\textbf {\bibinfo {volume}
  {2}},\ \bibinfo {pages} {4} (\bibinfo {year} {1990})},\ \Eprint
  {http://arxiv.org/abs/https://doi.org/10.1287/ijoc.2.1.4}
  {https://doi.org/10.1287/ijoc.2.1.4} \BibitemShut {NoStop}%
\bibitem [{\citenamefont {Fomin}\ \emph {et~al.}(2009)\citenamefont {Fomin},
  \citenamefont {Grandoni},\ and\ \citenamefont {Kratsch}}]{Fomin2009}%
  \BibitemOpen
  \bibfield  {author} {\bibinfo {author} {\bibfnamefont {F.~V.}\ \bibnamefont
  {Fomin}}, \bibinfo {author} {\bibfnamefont {F.}~\bibnamefont {Grandoni}}, \
  and\ \bibinfo {author} {\bibfnamefont {D.}~\bibnamefont {Kratsch}},\ }\href
  {\doibase 10.1145/1552285.1552286} {\bibfield  {journal} {\bibinfo  {journal}
  {Journal of the {ACM}}\ }\textbf {\bibinfo {volume} {56}},\ \bibinfo {pages}
  {1} (\bibinfo {year} {2009})}\BibitemShut {NoStop}%
\bibitem [{\citenamefont {van Rooij}\ \emph {et~al.}(2009)\citenamefont {van
  Rooij}, \citenamefont {Nederlof},\ and\ \citenamefont {van
  Dijk}}]{vanRooij2009}%
  \BibitemOpen
  \bibfield  {author} {\bibinfo {author} {\bibfnamefont {J.~M.~M.}\
  \bibnamefont {van Rooij}}, \bibinfo {author} {\bibfnamefont {J.}~\bibnamefont
  {Nederlof}}, \ and\ \bibinfo {author} {\bibfnamefont {T.~C.}\ \bibnamefont
  {van Dijk}},\ }in\ \href {\doibase 10.1007/978-3-642-04128-0_50} {\emph
  {\bibinfo {booktitle} {Lecture Notes in Computer Science}}}\ (\bibinfo
  {publisher} {Springer Berlin Heidelberg},\ \bibinfo {year} {2009})\ pp.\
  \bibinfo {pages} {554--565}\BibitemShut {NoStop}%
\bibitem [{\citenamefont {{Kadowaki}}\ and\ \citenamefont
  {{Nishimori}}(1998)}]{1998PhRvE..58.5355K}%
  \BibitemOpen
  \bibfield  {author} {\bibinfo {author} {\bibfnamefont {T.}~\bibnamefont
  {{Kadowaki}}}\ and\ \bibinfo {author} {\bibfnamefont {H.}~\bibnamefont
  {{Nishimori}}},\ }\href {\doibase 10.1103/PhysRevE.58.5355} {\bibfield
  {journal} {\bibinfo  {journal} {Phys. Rev. E}\ }\textbf {\bibinfo {volume}
  {58}},\ \bibinfo {pages} {5355} (\bibinfo {year} {1998})},\ \Eprint
  {http://arxiv.org/abs/cond-mat/9804280} {cond-mat/9804280} \BibitemShut
  {NoStop}%
\bibitem [{\citenamefont {{Farhi}}\ \emph {et~al.}(2000)\citenamefont
  {{Farhi}}, \citenamefont {{Goldstone}}, \citenamefont {{Gutmann}},\ and\
  \citenamefont {{Sipser}}}]{2000quant.ph..1106F}%
  \BibitemOpen
  \bibfield  {author} {\bibinfo {author} {\bibfnamefont {E.}~\bibnamefont
  {{Farhi}}}, \bibinfo {author} {\bibfnamefont {J.}~\bibnamefont
  {{Goldstone}}}, \bibinfo {author} {\bibfnamefont {S.}~\bibnamefont
  {{Gutmann}}}, \ and\ \bibinfo {author} {\bibfnamefont {M.}~\bibnamefont
  {{Sipser}}},\ }\href@noop {} {\bibfield  {journal} {\bibinfo  {journal}
  {eprint arXiv:quant-ph/0001106}\ } (\bibinfo {year} {2000})},\ \Eprint
  {http://arxiv.org/abs/quant-ph/0001106} {quant-ph/0001106} \BibitemShut
  {NoStop}%
\bibitem [{\citenamefont {Das}\ and\ \citenamefont
  {Chakrabarti}(2008)}]{RevModPhys.80.1061}%
  \BibitemOpen
  \bibfield  {author} {\bibinfo {author} {\bibfnamefont {A.}~\bibnamefont
  {Das}}\ and\ \bibinfo {author} {\bibfnamefont {B.~K.}\ \bibnamefont
  {Chakrabarti}},\ }\href {\doibase 10.1103/RevModPhys.80.1061} {\bibfield
  {journal} {\bibinfo  {journal} {Rev. Mod. Phys.}\ }\textbf {\bibinfo {volume}
  {80}},\ \bibinfo {pages} {1061} (\bibinfo {year} {2008})}\BibitemShut
  {NoStop}%
\bibitem [{\citenamefont {Takada}\ \emph {et~al.}(2020)\citenamefont {Takada},
  \citenamefont {Yamashiro},\ and\ \citenamefont
  {Nishimori}}]{doi:10.7566/JPSJ.89.044001}%
  \BibitemOpen
  \bibfield  {author} {\bibinfo {author} {\bibfnamefont {K.}~\bibnamefont
  {Takada}}, \bibinfo {author} {\bibfnamefont {Y.}~\bibnamefont {Yamashiro}}, \
  and\ \bibinfo {author} {\bibfnamefont {H.}~\bibnamefont {Nishimori}},\ }\href
  {\doibase 10.7566/JPSJ.89.044001} {\bibfield  {journal} {\bibinfo  {journal}
  {Journal of the Physical Society of Japan}\ }\textbf {\bibinfo {volume}
  {89}},\ \bibinfo {pages} {044001} (\bibinfo {year} {2020})},\ \Eprint
  {http://arxiv.org/abs/https://doi.org/10.7566/JPSJ.89.044001}
  {https://doi.org/10.7566/JPSJ.89.044001} \BibitemShut {NoStop}%
\bibitem [{\citenamefont {{Glover}}\ \emph {et~al.}(2018)\citenamefont
  {{Glover}}, \citenamefont {{Kochenberger}},\ and\ \citenamefont
  {{Du}}}]{2018Glover}%
  \BibitemOpen
  \bibfield  {author} {\bibinfo {author} {\bibfnamefont {F.}~\bibnamefont
  {{Glover}}}, \bibinfo {author} {\bibfnamefont {G.}~\bibnamefont
  {{Kochenberger}}}, \ and\ \bibinfo {author} {\bibfnamefont {Y.}~\bibnamefont
  {{Du}}},\ }\href@noop {} {\bibfield  {journal} {\bibinfo  {journal} {arXiv
  e-prints}\ ,\ \bibinfo {eid} {arXiv:1811.11538}} (\bibinfo {year} {2018})},\
  \Eprint {http://arxiv.org/abs/1811.11538} {arXiv:1811.11538 [cs.DS]}
  \BibitemShut {NoStop}%
\bibitem [{\citenamefont {Lanting}\ \emph {et~al.}(2017)\citenamefont
  {Lanting}, \citenamefont {King}, \citenamefont {Evert},\ and\ \citenamefont
  {Hoskinson}}]{PhysRevA.96.042322}%
  \BibitemOpen
  \bibfield  {author} {\bibinfo {author} {\bibfnamefont {T.}~\bibnamefont
  {Lanting}}, \bibinfo {author} {\bibfnamefont {A.~D.}\ \bibnamefont {King}},
  \bibinfo {author} {\bibfnamefont {B.}~\bibnamefont {Evert}}, \ and\ \bibinfo
  {author} {\bibfnamefont {E.}~\bibnamefont {Hoskinson}},\ }\href {\doibase
  10.1103/PhysRevA.96.042322} {\bibfield  {journal} {\bibinfo  {journal} {Phys.
  Rev. A}\ }\textbf {\bibinfo {volume} {96}},\ \bibinfo {pages} {042322}
  (\bibinfo {year} {2017})}\BibitemShut {NoStop}%
\bibitem [{\citenamefont {Hsu}\ \emph {et~al.}(2018)\citenamefont {Hsu},
  \citenamefont {Jin}, \citenamefont {Seidel}, \citenamefont {Neukart},
  \citenamefont {Raedt},\ and\ \citenamefont {Michielsen}}]{hsu2018quantum}%
  \BibitemOpen
  \bibfield  {author} {\bibinfo {author} {\bibfnamefont {T.-J.}\ \bibnamefont
  {Hsu}}, \bibinfo {author} {\bibfnamefont {F.}~\bibnamefont {Jin}}, \bibinfo
  {author} {\bibfnamefont {C.}~\bibnamefont {Seidel}}, \bibinfo {author}
  {\bibfnamefont {F.}~\bibnamefont {Neukart}}, \bibinfo {author} {\bibfnamefont
  {H.~D.}\ \bibnamefont {Raedt}}, \ and\ \bibinfo {author} {\bibfnamefont
  {K.}~\bibnamefont {Michielsen}},\ }\href@noop {} {\enquote {\bibinfo {title}
  {Quantum annealing with anneal path control: application to 2-sat problems
  with known energy landscapes},}\ } (\bibinfo {year} {2018}),\ \Eprint
  {http://arxiv.org/abs/1810.00194} {arXiv:1810.00194 [quant-ph]} \BibitemShut
  {NoStop}%
\bibitem [{\citenamefont {Yarkoni}\ \emph {et~al.}(2019)\citenamefont
  {Yarkoni}, \citenamefont {Wang}, \citenamefont {Plaat},\ and\ \citenamefont
  {B{\"a}ck}}]{10.1007/978-3-030-14082-3_14}%
  \BibitemOpen
  \bibfield  {author} {\bibinfo {author} {\bibfnamefont {S.}~\bibnamefont
  {Yarkoni}}, \bibinfo {author} {\bibfnamefont {H.}~\bibnamefont {Wang}},
  \bibinfo {author} {\bibfnamefont {A.}~\bibnamefont {Plaat}}, \ and\ \bibinfo
  {author} {\bibfnamefont {T.}~\bibnamefont {B{\"a}ck}},\ }in\ \href@noop {}
  {\emph {\bibinfo {booktitle} {Quantum Technology and Optimization
  Problems}}},\ \bibinfo {editor} {edited by\ \bibinfo {editor} {\bibfnamefont
  {S.}~\bibnamefont {Feld}}\ and\ \bibinfo {editor} {\bibfnamefont
  {C.}~\bibnamefont {Linnhoff-Popien}}}\ (\bibinfo  {publisher} {Springer
  International Publishing},\ \bibinfo {address} {Cham},\ \bibinfo {year}
  {2019})\ pp.\ \bibinfo {pages} {157--168}\BibitemShut {NoStop}%
\bibitem [{\citenamefont {Nandkishore}\ and\ \citenamefont
  {Huse}(2015)}]{doi:10.1146/annurev-conmatphys-031214-014726}%
  \BibitemOpen
  \bibfield  {author} {\bibinfo {author} {\bibfnamefont {R.}~\bibnamefont
  {Nandkishore}}\ and\ \bibinfo {author} {\bibfnamefont {D.~A.}\ \bibnamefont
  {Huse}},\ }\href {\doibase 10.1146/annurev-conmatphys-031214-014726}
  {\bibfield  {journal} {\bibinfo  {journal} {Annual Review of Condensed Matter
  Physics}\ }\textbf {\bibinfo {volume} {6}},\ \bibinfo {pages} {15} (\bibinfo
  {year} {2015})},\ \Eprint
  {http://arxiv.org/abs/https://doi.org/10.1146/annurev-conmatphys-031214-014726}
  {https://doi.org/10.1146/annurev-conmatphys-031214-014726} \BibitemShut
  {NoStop}%
\bibitem [{\citenamefont {Silaev}\ \emph {et~al.}(2014)\citenamefont {Silaev},
  \citenamefont {Heikkil\"a},\ and\ \citenamefont
  {Virtanen}}]{PhysRevE.90.022103}%
  \BibitemOpen
  \bibfield  {author} {\bibinfo {author} {\bibfnamefont {M.}~\bibnamefont
  {Silaev}}, \bibinfo {author} {\bibfnamefont {T.~T.}\ \bibnamefont
  {Heikkil\"a}}, \ and\ \bibinfo {author} {\bibfnamefont {P.}~\bibnamefont
  {Virtanen}},\ }\href {\doibase 10.1103/PhysRevE.90.022103} {\bibfield
  {journal} {\bibinfo  {journal} {Phys. Rev. E}\ }\textbf {\bibinfo {volume}
  {90}},\ \bibinfo {pages} {022103} (\bibinfo {year} {2014})}\BibitemShut
  {NoStop}%
\bibitem [{\citenamefont {Abanin}\ \emph {et~al.}(2019)\citenamefont {Abanin},
  \citenamefont {Altman}, \citenamefont {Bloch},\ and\ \citenamefont
  {Serbyn}}]{RevModPhys.91.021001}%
  \BibitemOpen
  \bibfield  {author} {\bibinfo {author} {\bibfnamefont {D.~A.}\ \bibnamefont
  {Abanin}}, \bibinfo {author} {\bibfnamefont {E.}~\bibnamefont {Altman}},
  \bibinfo {author} {\bibfnamefont {I.}~\bibnamefont {Bloch}}, \ and\ \bibinfo
  {author} {\bibfnamefont {M.}~\bibnamefont {Serbyn}},\ }\href {\doibase
  10.1103/RevModPhys.91.021001} {\bibfield  {journal} {\bibinfo  {journal}
  {Rev. Mod. Phys.}\ }\textbf {\bibinfo {volume} {91}},\ \bibinfo {pages}
  {021001} (\bibinfo {year} {2019})}\BibitemShut {NoStop}%
\bibitem [{\citenamefont {Alet}\ and\ \citenamefont
  {Laflorencie}(2018)}]{ALET2018498}%
  \BibitemOpen
  \bibfield  {author} {\bibinfo {author} {\bibfnamefont {F.}~\bibnamefont
  {Alet}}\ and\ \bibinfo {author} {\bibfnamefont {N.}~\bibnamefont
  {Laflorencie}},\ }\href {\doibase https://doi.org/10.1016/j.crhy.2018.03.003}
  {\bibfield  {journal} {\bibinfo  {journal} {Comptes Rendus Physique}\
  }\textbf {\bibinfo {volume} {19}},\ \bibinfo {pages} {498 } (\bibinfo {year}
  {2018})},\ \bibinfo {note} {quantum simulation / Simulation
  quantique}\BibitemShut {NoStop}%
\bibitem [{\citenamefont {Pal}\ and\ \citenamefont
  {Huse}(2010)}]{PhysRevB.82.174411}%
  \BibitemOpen
  \bibfield  {author} {\bibinfo {author} {\bibfnamefont {A.}~\bibnamefont
  {Pal}}\ and\ \bibinfo {author} {\bibfnamefont {D.~A.}\ \bibnamefont {Huse}},\
  }\href {\doibase 10.1103/PhysRevB.82.174411} {\bibfield  {journal} {\bibinfo
  {journal} {Phys. Rev. B}\ }\textbf {\bibinfo {volume} {82}},\ \bibinfo
  {pages} {174411} (\bibinfo {year} {2010})}\BibitemShut {NoStop}%
\bibitem [{\citenamefont {Bardarson}\ \emph {et~al.}(2012)\citenamefont
  {Bardarson}, \citenamefont {Pollmann},\ and\ \citenamefont
  {Moore}}]{PhysRevLett.109.017202}%
  \BibitemOpen
  \bibfield  {author} {\bibinfo {author} {\bibfnamefont {J.~H.}\ \bibnamefont
  {Bardarson}}, \bibinfo {author} {\bibfnamefont {F.}~\bibnamefont {Pollmann}},
  \ and\ \bibinfo {author} {\bibfnamefont {J.~E.}\ \bibnamefont {Moore}},\
  }\href {\doibase 10.1103/PhysRevLett.109.017202} {\bibfield  {journal}
  {\bibinfo  {journal} {Phys. Rev. Lett.}\ }\textbf {\bibinfo {volume} {109}},\
  \bibinfo {pages} {017202} (\bibinfo {year} {2012})}\BibitemShut {NoStop}%
\bibitem [{\citenamefont {Chang}\ \emph {et~al.}(2019)\citenamefont {Chang},
  \citenamefont {Gambhir}, \citenamefont {Humble},\ and\ \citenamefont
  {Sota}}]{Chang:2018uoc}%
  \BibitemOpen
  \bibfield  {author} {\bibinfo {author} {\bibfnamefont {C.~C.}\ \bibnamefont
  {Chang}}, \bibinfo {author} {\bibfnamefont {A.}~\bibnamefont {Gambhir}},
  \bibinfo {author} {\bibfnamefont {T.~S.}\ \bibnamefont {Humble}}, \ and\
  \bibinfo {author} {\bibfnamefont {S.}~\bibnamefont {Sota}},\ }\href {\doibase
  10.1038/s41598-019-46729-0} {\bibfield  {journal} {\bibinfo  {journal} {Sci.
  Rep.}\ }\textbf {\bibinfo {volume} {9}},\ \bibinfo {pages} {10258} (\bibinfo
  {year} {2019})},\ \Eprint {http://arxiv.org/abs/1812.06917} {arXiv:1812.06917
  [quant-ph]} \BibitemShut {NoStop}%
%%CITATION = ARXIV:1812.06917;%%
\bibitem [{\citenamefont {Esposito}\ \emph {et~al.}(2009)\citenamefont
  {Esposito}, \citenamefont {Harbola},\ and\ \citenamefont
  {Mukamel}}]{RevModPhys.81.1665}%
  \BibitemOpen
  \bibfield  {author} {\bibinfo {author} {\bibfnamefont {M.}~\bibnamefont
  {Esposito}}, \bibinfo {author} {\bibfnamefont {U.}~\bibnamefont {Harbola}}, \
  and\ \bibinfo {author} {\bibfnamefont {S.}~\bibnamefont {Mukamel}},\ }\href
  {\doibase 10.1103/RevModPhys.81.1665} {\bibfield  {journal} {\bibinfo
  {journal} {Rev. Mod. Phys.}\ }\textbf {\bibinfo {volume} {81}},\ \bibinfo
  {pages} {1665} (\bibinfo {year} {2009})}\BibitemShut {NoStop}%
\bibitem [{\citenamefont {Nielsen}\ and\ \citenamefont
  {Chuang}(2011)}]{10.5555/1972505}%
  \BibitemOpen
  \bibfield  {author} {\bibinfo {author} {\bibfnamefont {M.~A.}\ \bibnamefont
  {Nielsen}}\ and\ \bibinfo {author} {\bibfnamefont {I.~L.}\ \bibnamefont
  {Chuang}},\ }\href@noop {} {\emph {\bibinfo {title} {Quantum Computation and
  Quantum Information: 10th Anniversary Edition}}},\ \bibinfo {edition} {10th}\
  ed.\ (\bibinfo  {publisher} {Cambridge University Press},\ \bibinfo {address}
  {USA},\ \bibinfo {year} {2011})\BibitemShut {NoStop}%
\bibitem [{\citenamefont {Preskill}(1998)}]{preskill1998lecture}%
  \BibitemOpen
  \bibfield  {author} {\bibinfo {author} {\bibfnamefont {J.}~\bibnamefont
  {Preskill}},\ }\href@noop {} {\bibfield  {journal} {\bibinfo  {journal}
  {California Institute of Technology}\ }\textbf {\bibinfo {volume} {16}}
  (\bibinfo {year} {1998})}\BibitemShut {NoStop}%
\bibitem [{\citenamefont {{King}}\ \emph {et~al.}(2016)\citenamefont {{King}},
  \citenamefont {{Hoskinson}}, \citenamefont {{Lanting}}, \citenamefont
  {{Andriyash}},\ and\ \citenamefont {{Amin}}}]{2016PhRvA..93e2320K}%
  \BibitemOpen
  \bibfield  {author} {\bibinfo {author} {\bibfnamefont {A.~D.}\ \bibnamefont
  {{King}}}, \bibinfo {author} {\bibfnamefont {E.}~\bibnamefont {{Hoskinson}}},
  \bibinfo {author} {\bibfnamefont {T.}~\bibnamefont {{Lanting}}}, \bibinfo
  {author} {\bibfnamefont {E.}~\bibnamefont {{Andriyash}}}, \ and\ \bibinfo
  {author} {\bibfnamefont {M.~H.}\ \bibnamefont {{Amin}}},\ }\href {\doibase
  10.1103/PhysRevA.93.052320} {\bibfield  {journal} {\bibinfo  {journal}
  {\pra}\ }\textbf {\bibinfo {volume} {93}},\ \bibinfo {eid} {052320} (\bibinfo
  {year} {2016})},\ \Eprint {http://arxiv.org/abs/1512.07325} {arXiv:1512.07325
  [quant-ph]} \BibitemShut {NoStop}%
\bibitem [{\citenamefont {{Mandr{\`a}}}\ \emph {et~al.}(2017)\citenamefont
  {{Mandr{\`a}}}, \citenamefont {{Zhu}},\ and\ \citenamefont
  {{Katzgraber}}}]{2017PhRvL.118g0502M}%
  \BibitemOpen
  \bibfield  {author} {\bibinfo {author} {\bibfnamefont {S.}~\bibnamefont
  {{Mandr{\`a}}}}, \bibinfo {author} {\bibfnamefont {Z.}~\bibnamefont {{Zhu}}},
  \ and\ \bibinfo {author} {\bibfnamefont {H.~G.}\ \bibnamefont
  {{Katzgraber}}},\ }\href {\doibase 10.1103/PhysRevLett.118.070502} {\bibfield
   {journal} {\bibinfo  {journal} {\prl}\ }\textbf {\bibinfo {volume} {118}},\
  \bibinfo {eid} {070502} (\bibinfo {year} {2017})},\ \Eprint
  {http://arxiv.org/abs/1606.07146} {arXiv:1606.07146 [quant-ph]} \BibitemShut
  {NoStop}%
\bibitem [{\citenamefont {Systems}(2017)}]{dwave_temp}%
  \BibitemOpen
  \bibfield  {author} {\bibinfo {author} {\bibfnamefont {D.-W.}\ \bibnamefont
  {Systems}},\ }\href
  {https://www.dwavesys.com/sites/default/files/D-Wave%202000Q%20Tech%20Collateral_0117F_0.pdf}
  {\enquote {\bibinfo {title} {The d-wave 2000q quantum computer technology
  overview},}\ } (\bibinfo {year} {2017})\BibitemShut {NoStop}%
\bibitem [{\citenamefont {{Chiorescu}}\ \emph {et~al.}(2003)\citenamefont
  {{Chiorescu}}, \citenamefont {{Nakamura}}, \citenamefont {{Harmans}},\ and\
  \citenamefont {{Mooij}}}]{2003Sci...299.1869C}%
  \BibitemOpen
  \bibfield  {author} {\bibinfo {author} {\bibfnamefont {I.}~\bibnamefont
  {{Chiorescu}}}, \bibinfo {author} {\bibfnamefont {Y.}~\bibnamefont
  {{Nakamura}}}, \bibinfo {author} {\bibfnamefont {C.~J.~P.~M.}\ \bibnamefont
  {{Harmans}}}, \ and\ \bibinfo {author} {\bibfnamefont {J.~E.}\ \bibnamefont
  {{Mooij}}},\ }\href {\doibase 10.1126/science.1081045} {\bibfield  {journal}
  {\bibinfo  {journal} {Science}\ }\textbf {\bibinfo {volume} {299}},\ \bibinfo
  {pages} {1869} (\bibinfo {year} {2003})},\ \Eprint
  {http://arxiv.org/abs/cond-mat/0305461} {arXiv:cond-mat/0305461
  [cond-mat.mes-hall]} \BibitemShut {NoStop}%
\bibitem [{\citenamefont {{Luitz}}\ \emph {et~al.}(2015)\citenamefont
  {{Luitz}}, \citenamefont {{Laflorencie}},\ and\ \citenamefont
  {{Alet}}}]{2015PhRvB..91h1103L}%
  \BibitemOpen
  \bibfield  {author} {\bibinfo {author} {\bibfnamefont {D.~J.}\ \bibnamefont
  {{Luitz}}}, \bibinfo {author} {\bibfnamefont {N.}~\bibnamefont
  {{Laflorencie}}}, \ and\ \bibinfo {author} {\bibfnamefont {F.}~\bibnamefont
  {{Alet}}},\ }\href {\doibase 10.1103/PhysRevB.91.081103} {\bibfield
  {journal} {\bibinfo  {journal} {\prb}\ }\textbf {\bibinfo {volume} {91}},\
  \bibinfo {eid} {081103} (\bibinfo {year} {2015})},\ \Eprint
  {http://arxiv.org/abs/1411.0660} {arXiv:1411.0660 [cond-mat.dis-nn]}
  \BibitemShut {NoStop}%
\bibitem [{\citenamefont {Goldenfeld}(1992)}]{Goldenfeld:1992qy}%
  \BibitemOpen
  \bibfield  {author} {\bibinfo {author} {\bibfnamefont {N.}~\bibnamefont
  {Goldenfeld}},\ }\href@noop {} {\emph {\bibinfo {title} {{Lectures on phase
  transitions and the renormalization group}}}}\ (\bibinfo {year}
  {1992})\BibitemShut {NoStop}%
\bibitem [{\citenamefont {Systems}(2020{\natexlab{a}})}]{dwave_oceans}%
  \BibitemOpen
  \bibfield  {author} {\bibinfo {author} {\bibfnamefont {D.-W.}\ \bibnamefont
  {Systems}},\ }\href {https://github.com/dwavesystems/dwave-ocean-sdk}
  {\enquote {\bibinfo {title} {D-wave cloud client},}\ } (\bibinfo {year}
  {2020}{\natexlab{a}})\BibitemShut {NoStop}%
\bibitem [{\citenamefont {{Choi}}(2008)}]{2008arXiv0804.4884C}%
  \BibitemOpen
  \bibfield  {author} {\bibinfo {author} {\bibfnamefont {V.}~\bibnamefont
  {{Choi}}},\ }\href@noop {} {\bibfield  {journal} {\bibinfo  {journal} {ArXiv
  e-prints}\ } (\bibinfo {year} {2008})},\ \Eprint
  {http://arxiv.org/abs/0804.4884} {arXiv:0804.4884 [quant-ph]} \BibitemShut
  {NoStop}%
\bibitem [{\citenamefont {Systems}(2020{\natexlab{b}})}]{dwave_as}%
  \BibitemOpen
  \bibfield  {author} {\bibinfo {author} {\bibfnamefont {D.-W.}\ \bibnamefont
  {Systems}},\ }\href
  {https://support.dwavesys.com/hc/en-us/articles/360005267253-QPU-Specific-Anneal-Schedules}
  {\enquote {\bibinfo {title} {Qpu-specific anneal schedules},}\ } (\bibinfo
  {year} {2020}{\natexlab{b}})\BibitemShut {NoStop}%
\bibitem [{\citenamefont {Systems}(2020{\natexlab{c}})}]{dwave_as_docu}%
  \BibitemOpen
  \bibfield  {author} {\bibinfo {author} {\bibfnamefont {D.-W.}\ \bibnamefont
  {Systems}},\ }\href {https://docs.dwavesys.com/docs/latest/doc_qpu.html}
  {\enquote {\bibinfo {title} {Technical description of the d-wave quantum
  processing unit},}\ } (\bibinfo {year} {2020}{\natexlab{c}})\BibitemShut
  {NoStop}%
\bibitem [{\citenamefont {Chang}\ \emph
  {et~al.}(2020{\natexlab{a}})\citenamefont {Chang}, \citenamefont {Chen},\
  and\ \citenamefont {K\"o{}rber}}]{github:cchang5/quantum_linear_programming}%
  \BibitemOpen
  \bibfield  {author} {\bibinfo {author} {\bibfnamefont {C.~C.}\ \bibnamefont
  {Chang}}, \bibinfo {author} {\bibfnamefont {C.-C.}\ \bibnamefont {Chen}}, \
  and\ \bibinfo {author} {\bibfnamefont {C.}~\bibnamefont {K\"o{}rber}},\
  }\href@noop {} {\enquote {\bibinfo {title} {Quantum linear programming},}\
  }\bibinfo {howpublished}
  {\url{https://github.com/cchang5/quantum_linear_programming} \texttt{tag:
  arXiv}} (\bibinfo {year} {2020}{\natexlab{a}})\BibitemShut {NoStop}%
\bibitem [{\citenamefont {Chang}\ \emph
  {et~al.}(2020{\natexlab{b}})\citenamefont {Chang}, \citenamefont {K\"orber},\
  and\ \citenamefont {Walker-Loud}}]{Chang:2019khk}%
  \BibitemOpen
  \bibfield  {author} {\bibinfo {author} {\bibfnamefont {C.~C.}\ \bibnamefont
  {Chang}}, \bibinfo {author} {\bibfnamefont {C.}~\bibnamefont {K\"orber}}, \
  and\ \bibinfo {author} {\bibfnamefont {A.}~\bibnamefont {Walker-Loud}},\
  }\href {\doibase 10.21105/joss.02007} {\bibfield  {journal} {\bibinfo
  {journal} {J. Open Source Softw.}\ }\textbf {\bibinfo {volume} {5}},\
  \bibinfo {pages} {2007} (\bibinfo {year} {2020}{\natexlab{b}})},\ \Eprint
  {http://arxiv.org/abs/1912.03580} {arXiv:1912.03580 [hep-lat]} \BibitemShut
  {NoStop}%
\end{thebibliography}%
%========================================================================================

\end{document}